\documentclass[apj,iop]{emulateapj}
\usepackage{apjfonts}
\usepackage{xspace}

\newcommand{\jybm}{{\ensuremath{\rm Jy\,beam^{-1}}}\xspace}
\newcommand{\htw}{{\ensuremath{\rm H_2}}\xspace}
\newcommand{\ammo}{{\ensuremath{\rm NH_3}}\xspace}
\newcommand{\dammo}{{\ensuremath{\rm NH_2D}}\xspace}
\newcommand{\nh}{{\ensuremath{\rm N_2H^{+}}}\xspace}
\newcommand{\ndp}{{\ensuremath{\rm N_2D^{+}}}\xspace}

\newcommand{\ceio}{{\ensuremath{{\rm C^{18}O}}}\xspace}
\newcommand{\thco}{{\ensuremath{{\rm ^{13}CO}}}\xspace}
\newcommand{\co}{{\ensuremath{{\rm ^{12}CO}}}\xspace}
\newcommand{\source}{L1451-mm\xspace}

\newcommand{\rsun}{{\ensuremath{R_{\odot}}}\xspace}
\newcommand{\msun}{{\ensuremath{M_{\odot}}}\xspace}
\newcommand{\msunyr}{{\ensuremath{M_{\odot}\,{\rm yr^{-1}}}}\xspace}
\newcommand{\lsun}{{\ensuremath{L_{\odot}}}\xspace}
\newcommand{\kms}{{\ensuremath{{\rm km\, s^{-1}}}}\xspace}

\newcommand{\mstar}{M_*}
\newcommand{\mdisk}{M_{\rm disk}}
\newcommand{\rc}{R_{\rm c}}

\usepackage[stable]{footmisc}

\begin{document}
\title{The enigmatic core \source: a first hydrostatic core? or a hidden VeLLO?
\footnotemark[$\star$, $\star\star$, $\star\star\star$]}
\footnotetext[$\star$]{Based on observations carried out with the IRAM 30m Telescope. IRAM is supported by  INSU/CNRS (France), MPG (Germany) and IGN (Spain).}
\footnotetext[$\star\star$]{The Submillimeter Array is a joint project between the Smithsonian Astrophysical Observatory 
and the Academia Sinica Institute of Astronomy and Astrophysics and is funded by the 
Smithsonian Institution and the Academia Sinica.}
\footnotetext[$\star\star\star$]{Support for CARMA construction was derived from the states of 
California, Illinois, and Maryland, 
the James S. McDonnell Foundation, the Gordon and Betty Moore Foundation, the Kenneth T. 
and Eileen L. Norris Foundation, the University of Chicago, the Associates of the California Institute 
of Technology, and the National Science Foundation. Ongoing CARMA development and operations 
are supported by the National Science Foundation under a cooperative agreement, and by the 
CARMA partner universities.}

\shorttitle{A FHSC or a hidden VeLLO?}

\shortauthors{J. E. Pineda et al.}
\author{Jaime E. Pineda\altaffilmark{1,2}, 
H\'ector G. Arce\altaffilmark{3}, 
Scott Schnee\altaffilmark{4}, 
Alyssa A. Goodman\altaffilmark{1}, 
Tyler Bourke\altaffilmark{1}, 
Jonathan B. Foster\altaffilmark{1,5}, 
Thomas Robitaille\altaffilmark{1}, 
Joel Tanner\altaffilmark{3}, 
Jens Kauffmann\altaffilmark{1,6}, 
Mario Tafalla\altaffilmark{7}, 
Paola Caselli\altaffilmark{8}, 
Guillem Anglada\altaffilmark{9}
}
\altaffiltext{1}{Harvard-Smithsonian Center for Astrophysics, 60 Garden St., Cambridge, MA 02138, USA}
\altaffiltext{2}{Current address: ESO, Karl Schwarzschild Str. 2, 85748 Garching bei Munchen, Germany; 
and 
UK ALMA Regional Centre Node, Jodrell Bank Centre for Astrophysics, School of Physics and Astronomy, University of Manchester, Manchester, M13 9PL, UK}
\altaffiltext{3}{Department of Astronomy, Yale University, P.O. Box 208101, New Haven, CT 06520-8101, USA}
\altaffiltext{4}{National Radio Astronomy Observatory, 520 Edgemont Rd, Charlottesville, VA 22903, USA}
\altaffiltext{5}{Current address: Institute for Astrophysical Research, 725 Commonwealth Avenue, Boston, MA 02215, USA}
\altaffiltext{6}{Current address: NPP Fellow, Jet Propulsion Laboratory, 4800 Oak Grove Drive, Pasadena, CA 91109, USA}
\altaffiltext{7}{Observatorio Astron\'omico Nacional (IGN), Alfonso XII 3, E-28014 Madrid, Spain}
\altaffiltext{8}{School of Physics and Astronomy, University of Leeds, Leeds LS2 9JT, UK}
\altaffiltext{9}{Instituto de Astrof\'{\i}sica de Andaluc\'{\i}a, CSIC, Apartado 3004, E-18080 Granada, Spain}
\email{jaime.pineda@manchester.ac.uk}

\slugcomment{Accepted by ApJ, September 6, 2011}

\begin{abstract}
We present the detection of a dust continuum source at 3-mm (CARMA) and  
1.3-mm (SMA), and \co(2--1) emission (SMA) towards the  \source dense core. 
These detections suggest a compact object and an outflow where no point source 
at mid-infrared wavelengths is detected using Spitzer.
An upper limit for the dense core bolometric luminosity of 0.05\,\lsun is obtained.
By modeling the broadband SED and the continuum interferometric visibilities simultaneously, 
we confirm that a central source of heating is needed to explain the observations. 
This modeling also shows that the data can be well fitted by a 
dense core with a YSO and disk, or by a dense core with a central First Hydrostatic Core (FHSC). 
Unfortunately, we are not able to decide between these two models, which 
produce similar fits. 
We also detect \co(2--1) emission with red- and blue-shifted emission suggesting the 
presence of a slow and poorly collimated outflow, in opposition to what is 
usually found towards young stellar objects but in agreement with prediction from simulations of a FHSC. 
This presents the best candidate, so far, for a FHSC, an object 
that has been identified in simulations of collapsing dense cores.
Whatever the true nature of the central object in \source, 
this core presents an excellent laboratory to study the earliest 
phases of low-mass star formation.
\end{abstract}

\section{Introduction}
Star formation takes place in the densest regions of molecular clouds, 
usually referred to as dense cores. 
The parental molecular clouds show highly supersonic velocity dispersions, 
while the dense cores show subsonic levels of turbulence 
\citep{Goodman_1998-coherence,Caselli:2002-n2h+_maps}.
Recently, \cite{Pineda_2010-transition_coherence} showed that this 
transition in velocity dispersion is extremely sharp and it can be observed in \ammo(1,1) 
(see also Pineda et al. 2011, in prep).

Starless dense cores represent the initial conditions of star formation. 
\cite{Crapsi_2005-N2H_N2D} identify a sample of starless cores 
which show a number of signs indicating that they 
may be ``evolved'' and thus close to forming a star.

In the earliest phases of star formation a starless core undergoes a 
gravitational collapse. 
Increasing central densities will result in an increase in dust optical depth and thus 
cooling within the core will not be as efficient as in the earliest phases. 
This increases the gas temperature and generates more pressure. 
The first numerical simulation to study the formation of a protostar 
from an isothermal core \citep{Larson_1969-Collapse}, revealed the formation of a central 
adiabatic core, defined as a ``first hydrostatic core'' (hereafter FHSC).
This FHSC would then accrete more mass and undergo adiabatic contraction until \htw is 
dissociated, at which point it begins a second collapse until it forms  
a ``second hydrostatic core,'' which is the starting point for protostellar objects.

A few FHSC candidates have been suggested in the past.
\cite{Belloche_2006-Cha_MMS1} present single dish observations of the Cha-MMS1 dense core 
which combined with detections at 24 and 70\,$\mu$m with Spitzer suggest the presence 
of a first hydrostatic core or an extremely young protostar \citep[see also][]{Belloche_2011-LABOCA_Cha}.
\cite{Chen_2010-First_Core} present SMA observations of the continuum 
at 1.3-mm and \co(2--1) line in the L1448 region located in the Perseus cloud 
where no Spitzer (IRAC or MIPS) source is detected.
They detect a weak continuum source and a well collimated high-velocity 
outflow is observed in \co(2--1). 
\cite{Chen_2010-First_Core} analyze different scenarios to explain the observations 
and conclude that a FHSC provide the best case, however, 
no actual modeling of the interferometric observations is presented. 
Recently, \cite{Enoch_2010-FHSC_Perseus} present CARMA 3-mm continuum and 
deep Spitzer 70\,$\mu$m observations 
of another FHSC candidate (Per-Bolo~58) in the NGC1333 region also located in the Perseus cloud. 
In these observations they detect a weak source in the 3-mm continuum and 70\,$\mu$m. 
\cite{Enoch_2010-FHSC_Perseus} simultaneously modeled the broadband SED and 
the visibilities, allowing them to conclude that the best explanation for the central source 
is a FHSC. 
Dunham et al. (2011, submitted) present SMA 1.3-mm observations which reveal 
a collimated slow molecular outflow using \co(2--1) emission.

Another class of low luminosity objects has been identified thanks to Spitzer: 
Very Low Luminosity Objects \citep[VeLLOs, e.g.,][]{Young_2004-L1014_Vello,Bourke_2005-outflow_L1014,Dunham_2006-IRAM_VeLLO}, 
some of which are found within evolved cores \citep[as classified by][]{Crapsi_2005-N2H_N2D}. 
These objects have low intrinsic luminosities ($L<0.1~\lsun$) and are embedded in a dense core \citep{PPV:cores}. 
As VeLLOs have only recently been revealed by Spitzer \citep{Dunham_2008-VeLLO_survey}, 
it is not yet clear whether these are sub-stellar objects that are still forming, 
or low-mass protostars in a low-accretion state.

Broadband SED modeling of VeLLOs suggest that these sources can be explained 
as embedded YSOs with a surrounding disk. 
In the case of IRAM 04191+1522 (hereafter IRAM 04191), continuum observations 
using the IRAM Plateau de Bure interferometer (PdBI) were 
interpreted by \cite{Belloche_2002-Molecules_IRAM04191_PdBI} as produced 
from the dense core's inner part without the need for a disk. 

Recently, \cite{Maury_2010-PdBI_pilot_survey} presented high-resolution 
PdBI observations towards a sample of 5 Class 0 sources to study the 
binary fraction in the early stages of star formation. 
Their sample includes two previously known VeLLOs: L1521-F and IRAM 04191. 
Dust continuum emission is detected toward both objects, which may arise from 
either a circumstellar disk or from the inner parts of the envelope. Lack of 
detailed modeling of the SED or visibilities in these sources makes it hard 
to distinguish between these two scenarios.

This paper presents observations of \source, a low-mass 
core without any associated mid-infrared source in which we have detected 
compact thermal dust emission and a molecular outflow, 
along with models constructed  to derive the properties of this object. 
In \S\ref{sec:previous} we discuss previous observations of \source. 
In \S\ref{sec:observations} presents data used in this paper.
In \S\ref{sec:results} we present the analysis of the observations and 
radiative transfer models to reproduce the observed 
spectral energy distribution (SED) and the continuum visibilities to 
constrain the physical conditions of the source.
Finally, we present our conclusions in \S\ref{sec:discussion}.

\section{\source}\label{sec:previous}

\begin{figure*}
\plotone{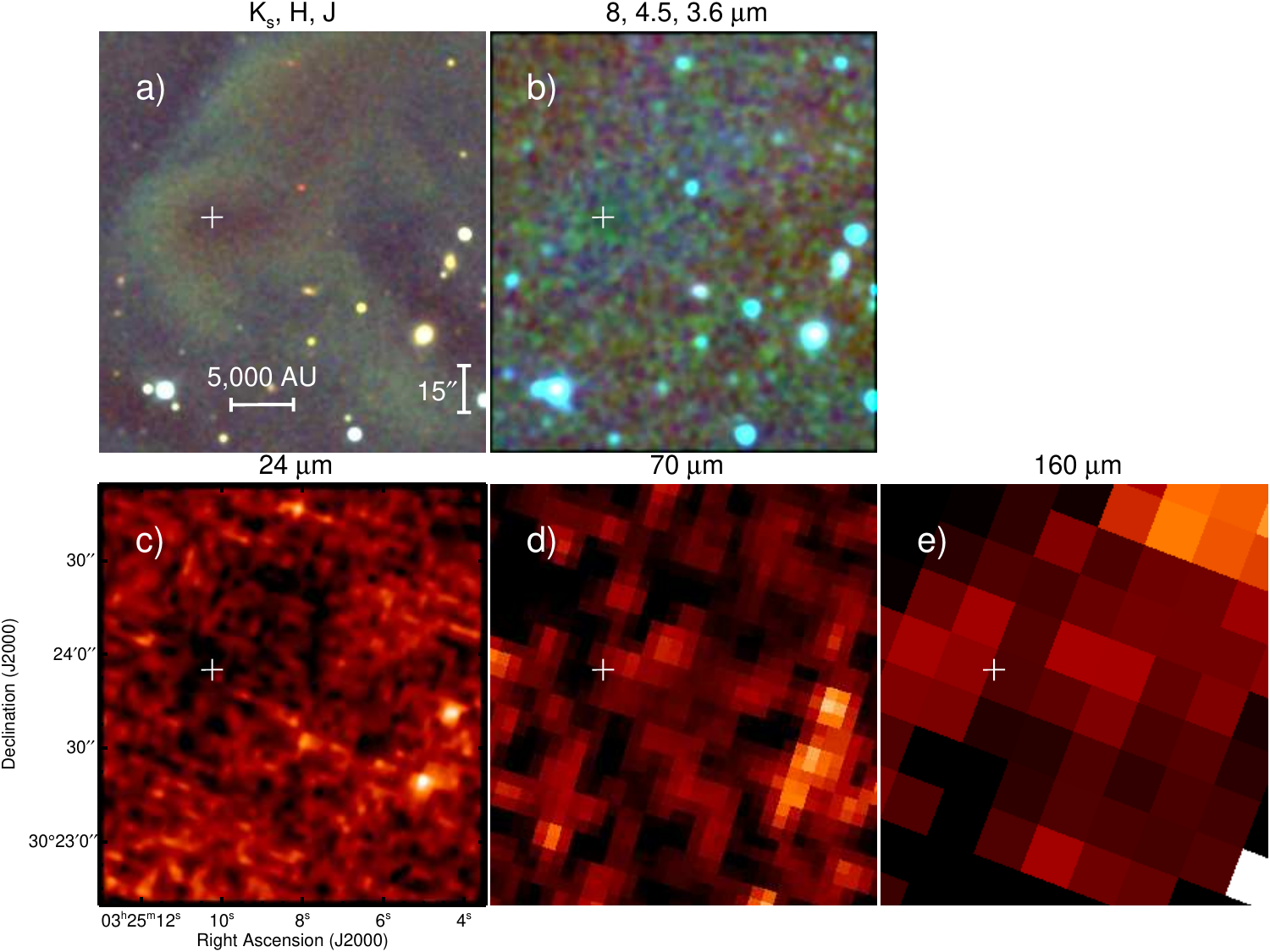}
\caption{Summary of observations available towards \source between 1 and 160$\mu$m.
{\bf (a) } $K_s$, $H$ and $J$ three color image, red, green, and blue respectively. 
The white cross shows the central position observed with the SMA. 
{\bf (b) } IRAC-4, -2 and -1 three color image, red, green, and blue respectively. 
{\bf (c)} MIPS-1 monochrome, 
{\bf (c)} MIPS-2 monochrome; and 
{\bf (d)} MIPS-3 monochrome figures. 
No point source is detected in any of the images.
\label{fig-images}}
\end{figure*}

\source \citep[also known as Per-Bolo~2;][]{Enoch_2006-Perseus_Bolocam} is a cold 
dense core in the L1451 Dark 
Cloud located in the Perseus Molecular Cloud Complex. 
Here we assume that Perseus is at a distance of $\sim 250$~pc \citep{Cernis_1990-Perseus_Distance,Hirota_2008-NGC1333_Distance}, 
which is consistent with those used by previous works. 
\source is detected in 1.1\,mm dust continuum with Bolocam at 31\arcsec\ resolution, 
and its estimated mass is 0.36~\msun from the Gaussian fit 
by \cite{Enoch_2006-Perseus_Bolocam} with major and minor FWHMs of $33\arcsec$ 
and $54\arcsec$, respectively. 
However, the core is too faint to be identified by the SCUBA surveys at 850~$\mu$m 
of the Perseus Cloud 
\citep{Hatchell_2005-SCUBA_Perseus,Kirk_2006-Perseus,Sadavoy_2010-Masses_GB}.

Figure~\ref{fig-images} presents a summary of the observations 
pre-dating this work towards \source. 
\cite{Foster_2006-cloudshine} presented deep Near-IR observations ($J$ $H$ $K_s$) of \source
which show only heavy obscuration, and no evidence for a point source. 
Establishing upper limits for this non-detection was complicated by the presence of 
extended bright structure (i.e., cloudshine) around the edge of \source. 
We estimate an upper limit by inserting synthetic stars with a range of magnitudes 
(in 0.1 magnitude steps) and appropriate FWHM at the central position. 
We ran Source Extractor \citep{Bertin_1996-SExtractor} on these synthetic images using a 
2.25$\arcsec$ radius aperture and established the input magnitude at which a 3$\sigma$ 
source was successfully extracted.

This core is classified as ``starless'' by \cite{Enoch_2008-starless_time}, because no point 
source is detected in Spitzer IRAC and MIPS images 
\citep{Jorgensen_2006-Perseus_IRAC,Rebull_2007-Perseus_MIPS}. 
Since the IRAC images do not contain significant extended emission we measured the flux 
in a 2.5$\arcsec$ radius aperture centered on the central position of \source with a background 
annulus of 2.5 to 7.5$\arcsec$ using the IRAF \verb+phot+ routine and applied the aperture 
correction factor for this configuration from the IRAC instrument handbook. 
All fluxes measured this way were within 2$\sigma$ of zero (fluxes were both positive and negative). 
For MIPS we used the smallest aperture with a well-defined aperture correction factor, 
which is 16$\arcsec$. 
Both MIPS1 and MIPS2 were consistent with zero flux while MIPS3 was a weak (2.7$\sigma$) detection. 
A summary of the photometric results is presented in Table~\ref{table:photo}.

\begin{deluxetable}{lccc}
\tablewidth{0pt}
\tablecaption{Photometry of \source \label{table:photo}}
\tablehead{%
\colhead{Filter}     &\colhead{Wavelength}   &           \colhead{Flux}             &  \colhead{Aperture} \\
\colhead{}        &    \colhead{($\mu$m)}&       \colhead{(mJy)}             &  \colhead{(arcsec)}
}
\startdata
$J$       & 1.25  &    $<0.006$      & 2.25\\
$H$      & 1.65  &    $<0.0048$    & 2.25\\
$K_s$  &  2.17 &    $<0.0099$    & 2.25\\
IRAC1  & 3.6    &    $<0.048$      &  2.5 \\
IRAC2  & 4.5    &    $<0.012$      & 2.5 \\
IRAC3  & 5.8    &    $<0.060$      &  2.5 \\
IRAC4  & 8.0    &    $<0.030$      &  2.5 \\
MIPS1  & 24.0  &    $<1.5$           & 16 \\
MIPS2  & 70.0  &    $<72$            & 16 \\
MIPS3  &160.   &   $880\pm330$          & 16 \\
IRAM    &1200  &   $70\pm 7$   & 16.8
\enddata
\tablecomments{Upper limits used are 3-$\sigma$ limits.}
\end{deluxetable}

Given the lack of detectable emission at Spitzer wavelengths, and using the 
correlation between 70$\mu$m and intrinsic YSO luminosity determined by 
\cite{Dunham_2008-VeLLO_survey}, 
an upper limit of $L < 1.6\times10^{-2}\, \lsun$ on the luminosity of a source 
embedded within \source  is determined.

For a given SED, two quantities can be calculated to describe it: 
bolometric luminosity, $L_{bol}$, and 
bolometric temperature, $T_{bol}$.  
The bolometric luminosity is calculated through integration of the SED ($S_{\nu}$) over 
the observed frequency range, 
\begin{equation}
L_{bol} = 4\pi \,d^{2} \int S_{\nu} d\nu~, 
\end{equation}
while the bolometric temperature is calculated following \cite{Myers_Ladd_1993-Tbol_Lbol}, 
\begin{equation}
T_{bol} = 1.25 \times 10^{-11} \left(\frac{\langle \nu \rangle}{{\rm Hz}}\right)\,{\rm K}~,
\end{equation}
where 
\begin{equation}
\langle \nu \rangle = \frac{\int \nu\,S_{\nu} d\nu}{\int S_{\nu} d\nu}~.
\end{equation}
For \source, if the upper limits are used as measurements, then we obtain 
$L_{bol} \le 0.05\,\lsun$ and $T_{bol} \le 30$\,K 
\citep[see][for discussions on the uncertainties in calculating $T_{bol}$ and $L_{bol}$]{Dunham_2008-VeLLO_survey,Enoch_2009-Properties_YSOs}.  
This bolometric luminosity is lower than any of the Class~0 objects studied by 
\cite{Enoch_2009-Properties_YSOs} in Serpens, Ophiuchus and 
Perseus Molecular Clouds; and also it is fainter than any of the VeLLOs 
with (sub-)millimeter wavelength observations studied by \cite{Dunham_2008-VeLLO_survey}.

Pineda et al. (2011 in prep) present \ammo(1,1) and (2,2) line maps 
observed with the 100-meter Green Bank Telescope.
From these observations they derive an almost constant (within a $\approx1\arcmin$ radius) 
kinetic temperature, $T_{kin}\approx 9.7$~K, and velocity dispersion, $\sigma_{v}\approx 0.15\,\kms$, 
showing no evidence for heating from a central source. 

\section{Observations}\label{sec:observations}

\subsection{Single dish continuum observations}
%

Dust continuum observations at 1.2-mm were taken 
using MAMBO at IRAM 30m telescope, under good weather ($\tau_{1.2mm}=0.1-0.2$). 
The data reduction was carried out using MOPSI, with  
parameters optimized for extended sources.
The observations are convolved with a 15\arcsec\ Gaussian kernel, while the flux 
unit is in Jy per 11\arcsec\ beam. 
The rms noise level is 1~mJy per 11\arcsec\ beam, and the map for the core studied is shown 
in Figure~\ref{fig_mambo_carma}.

\subsection{VLA Observations}

Observations were carried out with the Very Large Array (VLA) of the 
National Radio Astronomy Observatory on January 10, 2006 (project AA300).  
The \ammo$(J,K) = (1,1)$ and $(2,2)$ inversion transitions were observed simultaneously  
(see Table~\ref{sum-corr-vla} for a summary of the correlator configuration used). 
At this frequency the primary beam of the antennas is about $1.9\arcmin$.
The array was in the compact (D) configuration, the bandwidth was $1.56$ MHz, and 
the channel separation was 12.2~kHz (corresponding to 0.154\,\kms). 
This configuration is centered at the main hyperfine component and it also covers 
the inner pair of satellite lines for \ammo(1,1).

\begin{deluxetable}{ccccc}
\tablecolumns{5}
\tablewidth{0pt}
\tablecaption{VLA Spectral Setup: 23~GHz Setting
\label{sum-corr-vla}}
\tablehead{
\colhead{} & \colhead{} & 
\colhead{} & \colhead{Resolution} & 
\colhead{Frequency}\\
\colhead{Molecule} & \colhead{Transition} & 
\colhead{Chan.} & \colhead{(\kms)} & 
\colhead{(GHz)}
}
\startdata  
\ammo & $(1,1)$ & 63 & 0.1544 & 23.694495\\
\ammo & $(2,2)$ & 63 & 0.1542 & 23.722733
\enddata
\end{deluxetable}

The bandpass and absolute flux calibrator was the quasar 0319+415 (3C84) 
with a calculated flux density of 
$10.6$~Jy at $1$~cm, and the phase and amplitude calibrator was 0336+323.  
The raw-data were reduced using CASA image processing software. 
The signal from each baseline was inspected, and baselines showing spurious data were 
removed prior to imaging. 
The images were created using multi-scale clean (scales [8,24,72]~arcsec and smallscalebias=0.8) 
with a robust parameter of $0.5$ and tapering the image with a 8$\arcsec$ Gaussian to increase 
the signal-to-noise. 
Each channel was cleaned separately according to the spatial distribution of the emission, 
using a circular beam of $8\arcsec$.
Table~\ref{sum-map-beams} lists relevant information on the maps used.

\begin{deluxetable*}{lccc}
\tablecolumns{4}
\tablewidth{0pt}
\tablecaption{Parameters of Interferometric Maps
\label{sum-map-beams}}
\tablehead{
\colhead{Map} & \colhead{Array} & 
\colhead{Beam\tablenotemark{a}} &
\colhead{rms}
}
\startdata
\ammo(1,1) & VLA & 8\arcsec$\times$8\arcsec ($0\degr$) & 3\,m\jybm\,channel$^{-1}$\\
\ammo(2,2) & VLA & 8\arcsec$\times$8\arcsec ($0\degr$) & 3\,m\jybm\,channel$^{-1}$\\
\dammo($1_{11}$--$1_{01}$) & CARMA & 9.2\arcsec$\times$7.6\arcsec ($+72.3\degr$) & 90\,m\jybm\,channel$^{-1}$\\
\nh(1--0)   & CARMA & 5.2\arcsec$\times$4.3\arcsec ($-73\degr$) & 80\,m\jybm\,channel$^{-1}$\\
3-mm continuum & CARMA & 5.4\arcsec$\times$4.8\arcsec ($-77\degr$) & 0.5\,m\jybm\\ 
\co(2--1) & SMA & 1.35\arcsec$\times$0.96\arcsec ($+80.9\degr$) & 40\,m\jybm\,channel$^{-1}$\\
1.3-mm continuum & SMA & 1.23\arcsec$\times$0.88\arcsec ($+85.6\degr$) & 0.5\,m\jybm
\enddata
\tablenotetext{a}{Size and position angle. Position angle is measured counter clockwise from north.}
\end{deluxetable*}

\subsection{CARMA observations} 
Continuum observations in the 3-mm window were obtained with CARMA, a
15-element interferometer consisting of nine 6.1-meter antennas and
six 10.4-meter antennas, between April and September 2008.  
The CARMA correlator records signals in
three separate bands, each with an upper and lower sideband.  We
configured one band for maximum bandwidth (468\,MHz with 15 channels)
to observe continuum emission, providing a total continuum bandwidth
of 936\,MHz.
The remaining two bands were configured for maximum
spectral resolution (1.92\,MHz per band) to
observe \dammo($1_{11}$--$1_{01}$) and \nh(1--0)  
(see Table~\ref{sum-corr-carma} for the correlator configuration summary). 
The six
main hyperfine components of \dammo fit in the two narrow spectral
bands and six of the seven-hyperfine components of \nh(1--0)
were observed, with the highest frequency (isolated) component falling
outside the observed frequency range.

The field of view (half-power beam width) of the 10.4-m antennas is
$66\arcsec$ at the observed frequencies.  Seven point mosaics were made
around the center of \source in CARMA's D and E-array configurations,
giving baselines that range from 8-m to 150-m.  Observations of
\nh (but not \dammo) were also made in CARMA's C-array
configuration, with projected baselines of 30-m to 350-m.  The
synthesized beam sizes and position angles (measured counter clockwise from North) are:
5.4\arcsec$\times$4.8\arcsec\ and $-$77$\degr$ (continuum), 
5.2\arcsec$\times$4.3\arcsec\ and $-$73$\degr$ (\nh),
9.2\arcsec$\times$7.6\arcsec\ and 72.3$\degr$ (\dammo).  The largest
angular size to which these observations were sensitive is
$\sim$40\arcsec.

\begin{deluxetable}{lccccc}
\tablecolumns{6}
\tablewidth{0pt}
\tablecaption{CARMA Spectral Setup: 3-mm Setting
\label{sum-corr-carma}}
\tablehead{
\colhead{} & \colhead{} & 
\colhead{} & 
\colhead{} & \colhead{Resolution} & 
\colhead{Frequency}\\
\colhead{Molecule} & \colhead{Transition} & 
\colhead{Sideband} & 
\colhead{Chan.} & \colhead{(\kms)} & 
\colhead{(GHz)}
}
\startdata  
\dammo &$1_{11}$--$1_{01}$       & Lower & 2$\times$63 & 0.106 & 85.9262\\
\nh 	      &1--0                                    & Upper & 2$\times$63 & 0.098 & 93.1737
\enddata
\end{deluxetable}

The observing sequence for the CARMA observations was to integrate on
a primary and secondary phase calibrator (3C~111 and 0336+323) for 3
minutes each and the science target for 14 minutes.  In each set of
observations 3C~111 was used for passband calibration and observations
of Uranus were used for absolute flux calibration.  Based on the
repeatability of the quasar fluxes, the estimated random uncertainty
in the measured fluxes is $\sigma\simeq5$\%.  Radio pointing was done
at the beginning of each track and pointing constants were updated at
least every two hours thereafter, using either radio or optical
pointing routines \citep{Corder_2010-CARMA}.  
Calibration and imaging were done using the MIRIAD
data reduction package \citep{Sault95}.
Table~\ref{sum-map-beams} lists relevant information on the maps used.

\subsection{SMA observations}
The SMA observations were carried out at 1.3-mm (230~GHz) in both compact and extended configuration.
The compact array observations were carried out on November 1, 2009, 
with zenith opacity at 225~GHz of $\sim$0.085. 
Quasars 3C~84 and 3C~111 were observed for gain calibration.
Flux calibration was done with observations of Uranus and Ganymede.
Bandpass calibration was done using observations of the quasar 3C~273. 
The SMA correlator covers 2~GHz bandwidth in each of the
two sidebands. Each band is divided into
24 ``chunks'' of 104 MHz width, which can be covered by varying
spectral resolution.
The correlator configuration is summarized in Table~\ref{sum-corr-comp}.

The extended array observations were carried out on September 13, 2010, 
with zenith opacity at 225~GHz of $\sim$0.05. 
Quasars 3C~84 and 3C~111 were observed for gain calibration.
Flux calibration was done with observations of Uranus and Callisto.
Bandpass calibration was done using observations of the quasar 3C~454.3.
The SMA correlator used the new 4~GHz bandwidth in each of the
two sidebands. Each band is divided into
48 ``chunks'' of 104 MHz width, which can be covered by varying
spectral resolution.
The correlator configuration is summarized in Table~\ref{sum-corr-ext}. 

Both data sets were edited and calibrated using the MIR software 
package\footnote{See \url{http://cfa-www.harvard.edu/$\sim$cqi/mircook.html.}} adapted for the SMA. 
Imaging was performed with
the MIRIAD package \citep{Sault95}, resulting in an angular resolution of
1.23\arcsec$\times$0.88$\arcsec$ PA=85.6$\degr$ (using robust weighting parameter of -2) 
and  
1.35\arcsec$\times$0.96$\arcsec$ PA=80.9$\degr$  (using robust weighting parameter of 0) 
for the continuum and \co(2--1), respectively. 
Table~\ref{sum-map-beams} lists relevant information on the maps used.
The rms sensitivity is $\approx$0.5\,m\jybm 
for the continuum, using both sidebands (avoiding the chunk  
containing the \co line), and $\sim$36\,m\jybm per channel for
the line \co(2--1) data. 
The primary beam FWHM of the SMA at these frequencies is about 
55$\arcsec$.

\begin{deluxetable}{lccccc}
\tablecolumns{6}
\tablewidth{0pt}
\tablecaption{SMA (compact configuration) Spectral Setup: 1.3-mm Setting
\label{sum-corr-comp}}
\tablehead{
\colhead{} & \colhead{} & 
\colhead{} & 
\colhead{} & \colhead{Resolution} & 
\colhead{Frequency}\\
\colhead{Molecule} & \colhead{Transition} & 
\colhead{Chunk} & 
\colhead{Chan.} & \colhead{(\kms)} & 
\colhead{(GHz)}
}
\startdata
\cutinhead{LSB}
\ceio      & 2--1        &  s23 & 512 & 0.28 & 219.560357\\
\thco     & 2--1         &  s13 & 256 & 0.55 & 220.398684\\
\cutinhead{USB}
\co         &  2--1        &  s14 & 256 & 0.53 & 230.537964\\
\ndp       &  3--2        & s23 & 512 & 0.26 & 231.321966
\enddata
\tablecomments{For all other chunks the channels have a resolution of 0.8125~MHz.}
\end{deluxetable}

\begin{deluxetable}{lccccc}
\tablecolumns{6}
\tablewidth{0pt}
\tablecaption{SMA (extended configuration) Spectral Setup: 1.3-mm Setting
\label{sum-corr-ext}}
\tablehead{
\colhead{} & \colhead{} & 
\colhead{} & 
\colhead{} & \colhead{Resolution} & 
\colhead{Frequency}\\
\colhead{Molecule} & \colhead{Transition} & 
\colhead{Chunk} & 
\colhead{Chan.} & \colhead{(\kms)} & 
\colhead{(GHz)}
}
\startdata
\cutinhead{LSB}
\ceio      & 2--1        &  s23 & 128 & 1.11 & 219.560357\\
\thco     & 2--1         &  s13 & 128 & 1.11 & 220.398684\\
\cutinhead{USB}
\co         &  2--1        &  s14 & 256 & 0.53 & 230.537964\\
\ndp       &  3--2        & s23 & 128 & 1.05 &231.321966
\enddata
\tablecomments{The channels of all other chunks have resolution 0.8125~MHz, except those in 
chunks s15 and s16 where the resolution is 1.625~MHz.}
\end{deluxetable}

\section{Results}\label{sec:results}
\subsection{MAMBO and CARMA continuum}

\begin{figure*}
\plotone{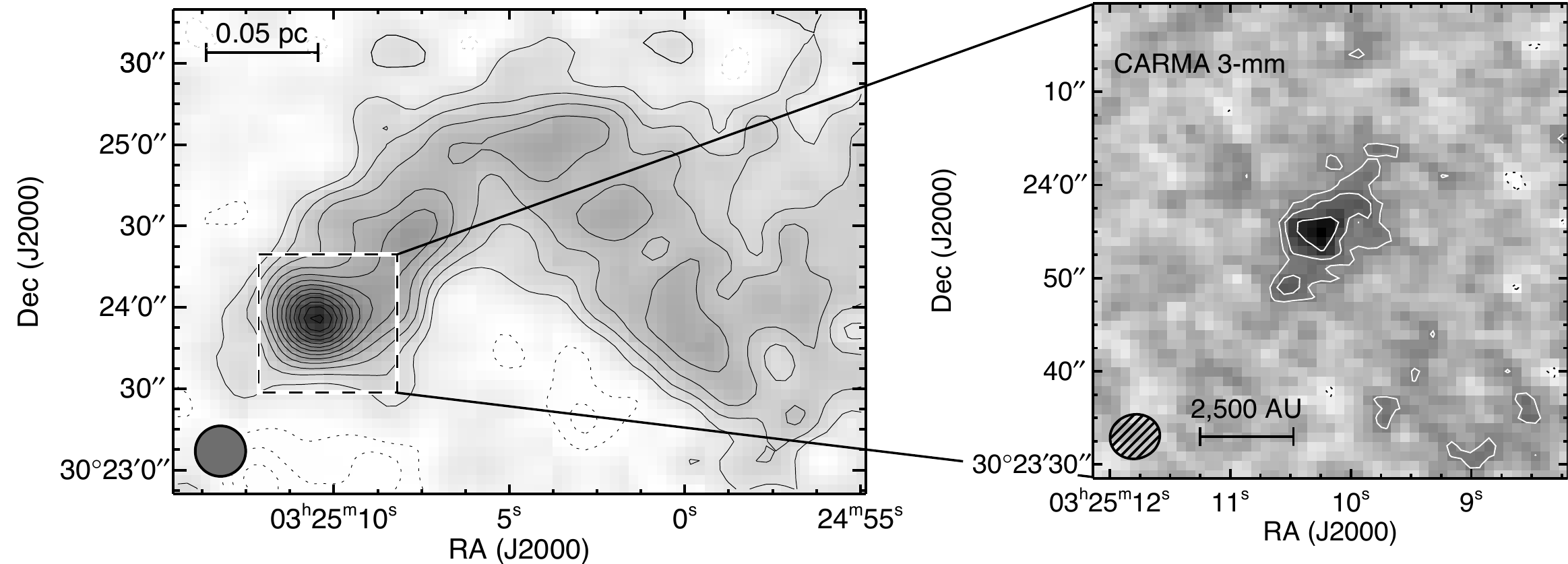}
\caption{Left panel shows the MAMBO 1.2~mm dust continuum emission, where there is both 
a compact central bright object 
($S_{peak}=33~$m\jybm) and also less bright and diffuse emission. 
Black solid contours represent [1, 2, 3, \ldots, 15]$\times$2.5~m\jybm levels, 
and dashed contours are -2.5 and -5~m\jybm levels. 
The dashed rectangle shows the area imaged by CARMA data. 
Right panel shows the gray scale 3-mm continuum emission map observed with CARMA. 
Solid contours mark the [2, 3, 5]$\times$0.5\,m\jybm, while negative contours are 
shown by dotted lines. 
A faint central source is detected at the center of the image that matches the 
MAMBO peak position.
\label{fig_mambo_carma}}
\end{figure*}

\begin{figure}
\plotone{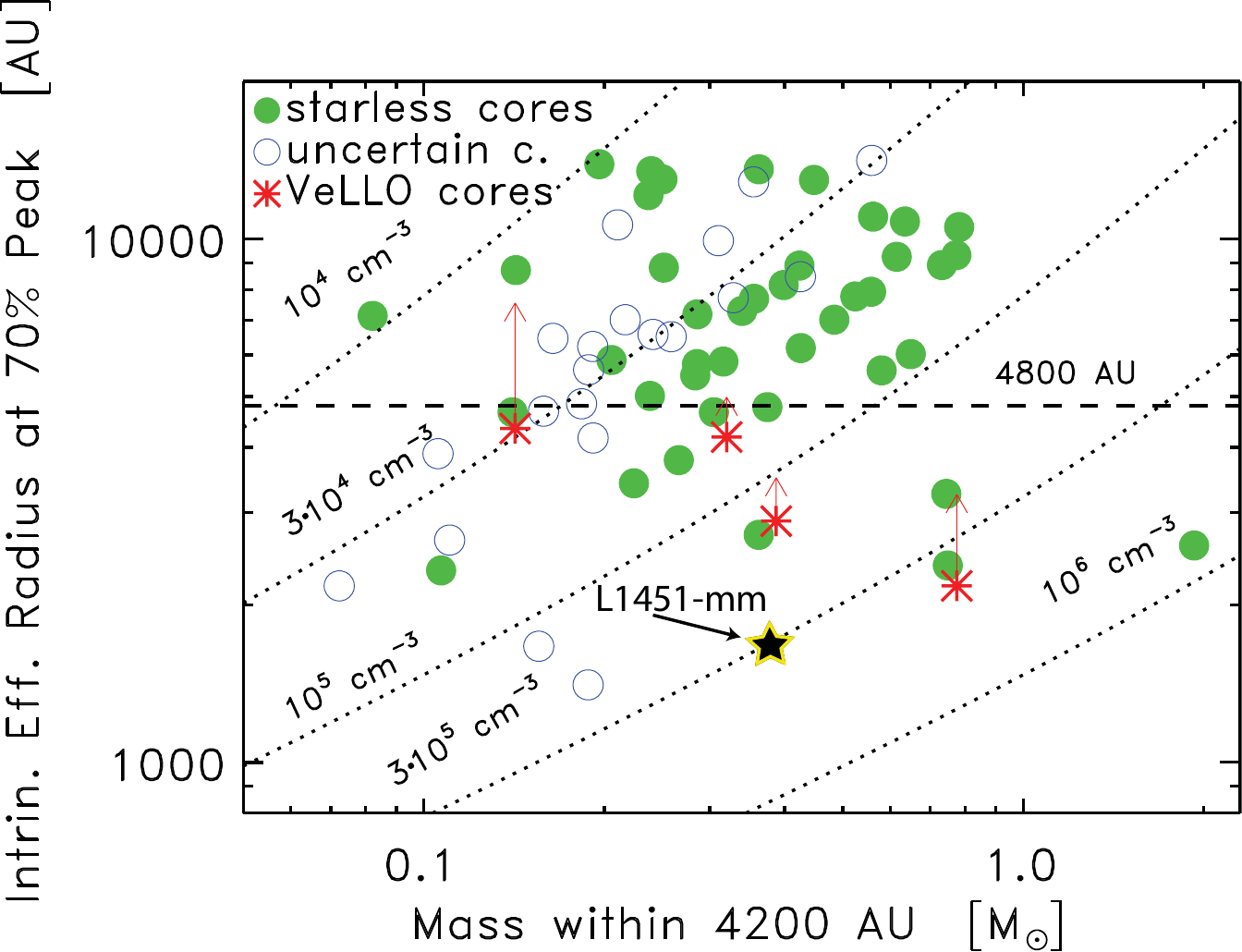}
\caption{Relation between the intrinsic radius at 70\% peak intensity and
the mass within 4200~AU radius from the peak for \source (show by the solid star), 
compared to the sample of starless cores and VeLLOs candidates 
from \cite{Kauffmann_2008-MAMBO_c2d}. 
Both properties (mass and radius) are derived from MAMBO data.
Starless cores with well established and uncertain properties are shown by 
filled and open circles, respectively. 
Cores hosting candidate VeLLOs are shown by stars, the radius bias 
due to internal heating by the central object is indicated by the arrows 
\citep[see][for details]{Kauffmann_2008-MAMBO_c2d}.
Curves of constant \htw central density are shown by dotted lines.
The dashed line
indicates the upper radius limit for evolved dense cores, $\le$4800~AU,
suggested by \cite{Crapsi_2005-N2H_N2D}. 
\label{fig_mambo}}
\end{figure}

The MAMBO dust continuum emission map (left panel of Figure~\ref{fig_mambo_carma})  
can be decomposed into a bright compact core, and fainter filamentary emission. 
The compact core peak position is located at $(\alpha,\delta)=$(03:25:10.4, +30:23:56.0). 
The compact core within 4,200~AU mass is estimated to be 0.3\,\msun, where a 
dust opacity per dust mass of 1.14\,cm$^{2}$\,g$^{-1}$ \citep{OH94}, gas-to-dust ratio of 100, 
and a dust temperature of 10~K are used. 
The MAMBO derived mass is consistent with the mass previously estimated using Bolocam.
In Figure~\ref{fig_mambo} the compact core is compared to the sample of 
starless cores from \cite{Kauffmann_2008-MAMBO_c2d}, where the fiducial radius 
of $4,200$~AU is used to compare with previous works \citep[e.g.,][]{Motte_2001-MAMBO_Survey}. 
suggesting that it is more compact than most starless cores
From this comparison we can establish that 
\source  is in fact quite compact, and therefore dense, 
suggesting that it is more compact than most starless cores
an evolved evolutionary state \citep[see][]{Crapsi_2005-N2H_N2D}.

A faint central source is detected in the CARMA 3-mm continuum map shown 
in right panel of Figure~\ref{fig_mambo_carma}. 
The continuum emission map is fitted by a Gaussian with a total flux of 10\,mJy, 
while if a point source is fitted a flux of (4$\pm$2)\,mJy is obtained. 
A summary of the fits to the CARMA 3-mm continuum are listed in Table~\ref{tab-carma-fit}. 
The CARMA continuum emission agrees with the MAMBO peak position.

\begin{deluxetable*}{lccccc}[h]
\tablecolumns{6}
\tablecaption{Results of Fits to CARMA 3-mm continuum image for L1451-mm
\label{tab-carma-fit}}
\tablewidth{0pt}
\tablehead{
\colhead{}              & \multicolumn{2}{c}{Center\tablenotemark{a}} & \colhead{Peak Flux} & \colhead{Size (FWHM)} & \colhead{PA}\\
\colhead{Source} & \colhead{$\alpha$(J2000)} & \colhead{$\delta$(J2000)} & \colhead{(mJy)} & \colhead{(arcsec)} & \colhead{($\deg$)}
}
\startdata
Point Source & 3:25:10.25 & +30:23:55.09 & 4$\pm$2 & \nodata & \nodata \\
Gaussian       & 3:25:10.21 & +30:23:55.20 & 2$\pm$1 & (16$\pm$13, 8$\pm$6) & $-$-37$\pm$44
\enddata
\tablenotetext{a}{Units of R.A. are hours, minutes, and seconds. 
Units of declination are degrees, arcminutes, and arcseconds.}
\end{deluxetable*}

\subsection{Molecular lines with VLA and CARMA}

\begin{figure*}
\plotone{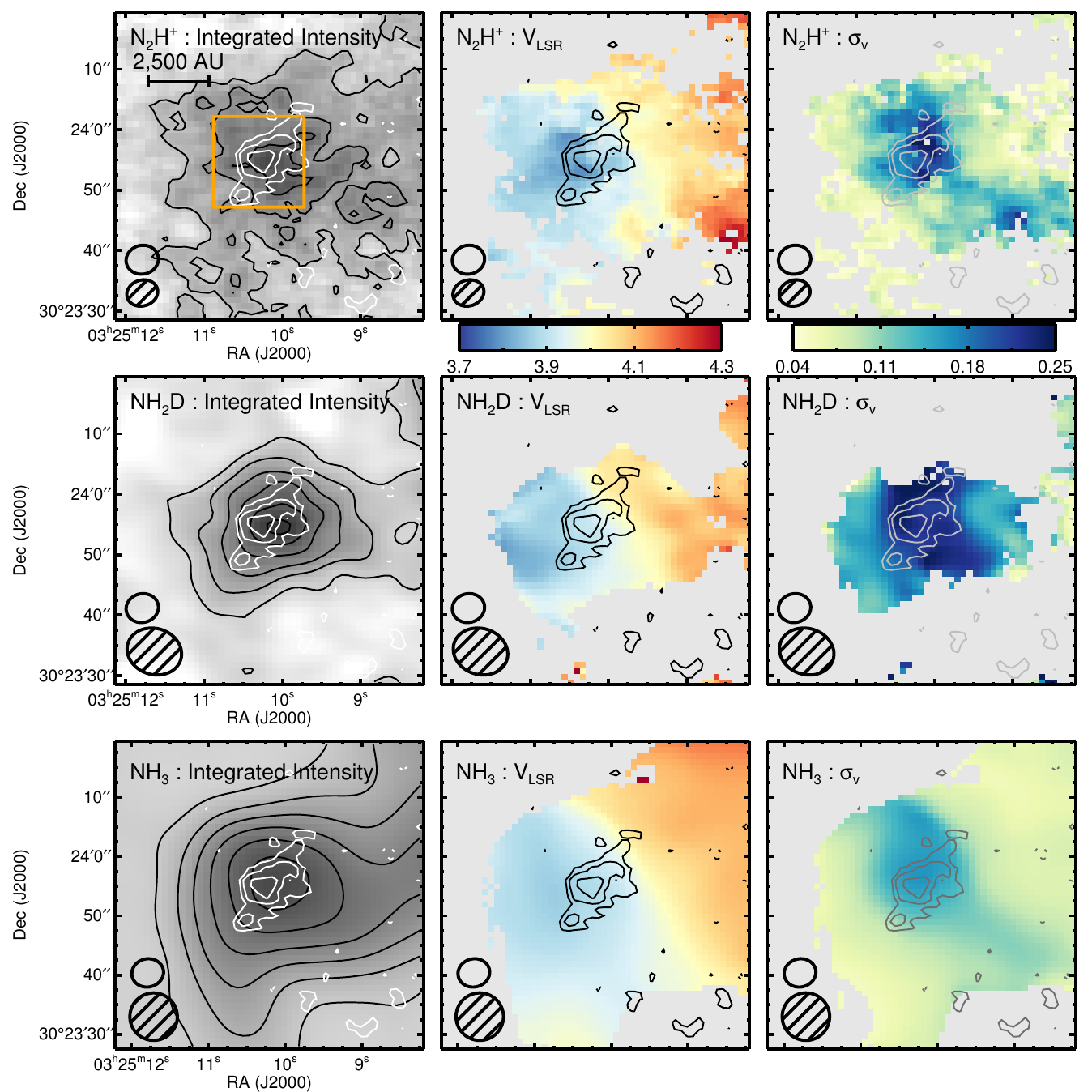}
\caption{Top, middle, and bottom rows present results from \nh (CARMA), \dammo (CARMA), 
and \ammo (VLA) emission line maps, respectively, where all observed hyperfine components are used. 
Left, middle, and right columns show the integrated intensity, centroid velocity, 
and velocity dispersion maps, respectively. 
Left panels also show contours for the integrated intensity at the following levels: 
[5,15,25]\,\jybm\kms for \nh (top), 
[5,15,25,35,45,55]\,\jybm\kms for \dammo (middle), 
[12.5,50,87.5,125,162.5,200]\,m\jybm\kms for \ammo (bottom). 
The color scales for the velocity (centroid and dispersion) maps are in \kms. 
Contours show the CARMA 3-mm continuum emission presented in Figure~\ref{fig_mambo_carma}. 
The synthesized beam for each transition line is shown at the bottom left corner and 
the CARMA 3-mm continuum beam shown above. 
The orange box in the upper left panel shows the region imaged using the SMA.
\label{fig-carma-mol}}
\end{figure*}

Figure~\ref{fig-carma-mol} shows the summary of the molecular line transitions  
observed with CARMA and VLA. 
In this study we will briefly discuss the kinematics of the region 
and leave a more in depth study of the core in forthcoming papers 
(Schnee et al., 2011 in prep., Arce et al., 2011 in prep.).
From these observations a centroid velocity and velocity dispersion are obtained by fitting the 
line profiles \citep[see][for details]{GBT:Perseus}.
The integrated intensity maps 
show the extended emission from the core where the peaks match the 
position of the CARMA continuum emission to within the respective beam size. 
The centroid velocity maps, for all three lines, show a consistent result with a clear 
velocity gradient. 
A gradient is fitted to the centroid velocity map for all three lines, with an average value of 
$\mathcal{G}=$6.1\,km\,s$^{-1}$\,pc$^{-1}$ and a $-66\degr$ position angle (measured counter 
clockwise from north), see Table~\ref{table-grad} for the individual fit obtained for all three maps. 
This velocity gradient is larger than those observed in lower angular resolution \ammo(1,1) 
maps \citep{Alyssa_1993-gradients} or using lower density tracers \citep{Kirk_2010-Core_Dynamics}, 
while velocity gradients of a similar magnitude are obtained with high angular resolution observations 
of dense gas \citep{Curtis_2011-Perseus_Core_kinematics,Tanner_2011-VLA_HH211}.
%
The velocity dispersion maps show 
a clear increase towards the center of the map, starting with very narrow lines 
(close to the thermal values) in the outer regions.  
The increase in velocity dispersion is more 
pronounced in the \nh velocity dispersion map, and the difference can be 
explained by the higher angular resolution obtained in the \nh observations.

\begin{deluxetable}{lccc}[h]
\tablewidth{0pt}
\tablecaption{Velocity Gradients Fits\label{table-grad}}
\tablehead{%
\colhead{Transition} & \colhead{$\mathcal{G}$} & \colhead{PA} & \colhead{$v_{0}$}\\
\colhead{} & \colhead{(km\,s$^{-1}$\,pc$^{-1}$)} & \colhead{(deg)} & \colhead{(\kms)}
}
\startdata
\nh         & $5.6\pm 0.2$     & $-84\pm  3$      & $3.970  \pm  0.002$\\
\dammo& $8\pm 1$     & $-83   \pm 15$        & $3.949   \pm 0.009$\\
\ammo  & $6.24\pm 0.06$ & $-65.8\pm0.7$ & $4.0042\pm  0.0006$
\enddata
\end{deluxetable}

\begin{figure*}
\plotone{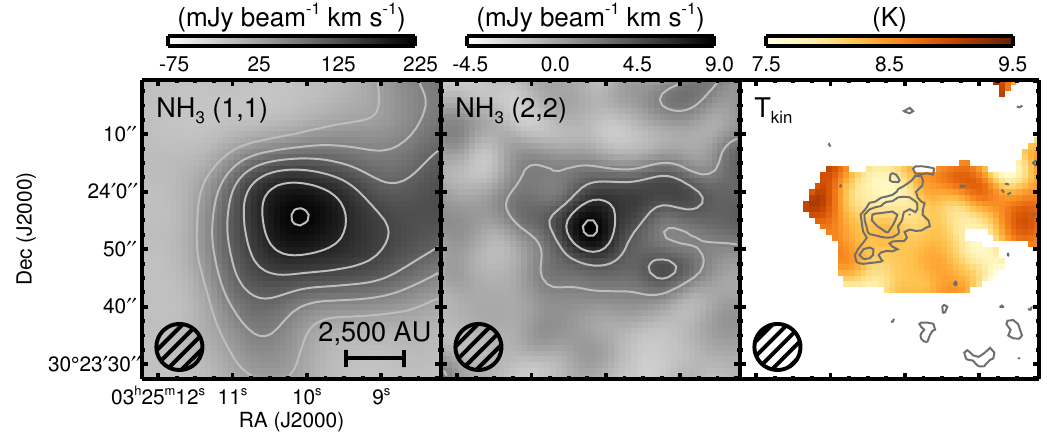}
\caption{Left and middle panels show the \ammo(1,1) and (2,2) integrated 
intensity VLA maps of the same region shown in Figure~\ref{fig-carma-mol} 
using all the observed hyperfine components. 
The overlaid contours show the integrated intensity at 
[2,4,6,\ldots,12] $\times$ 18.12 and 1.09\,m\jybm\kms for \ammo(1,1) and (2,2) in the 
left and middle panels, respectively. 
The right panel shows the kinetic temperature derived by fitting 
simultaneously both \ammo lines, overlaid with the CARMA 3-mm continuum. 
The kinetic temperature uncertainties ranges between 0.2\,K in the central region up to 0.5\,K 
in the outer region.
The kinetic temperature map presents small variations, where there is no increase 
in temperature at the peak position of the continuum emission. 
\label{fig-vla-temp}}
\end{figure*}

The \ammo(1,1) and (2,2) integrated intensity maps obtained using the VLA 
are shown in Figure~\ref{fig-vla-temp}, left and middle panels respectively, 
where all components observed are taken into account. 
Both lines present a peak coincident with the CARMA continuum peak, 
and where the \ammo(1,1) emission covers a more extended region 
than the \ammo(2,2). 
However, since the emission \ammo(1,1) is fairly extended, the addition of GBT 
data to provide the zero-spacing is needed to allow a robust temperature 
determination.
The morphology of the \ammo and \nh integrated intensity maps show non-flattened structures,
which are drastically different from those seen in young Class~0 sources 
\citep[e.g.,][]{Wiseman_2001-HH212_NH3,Chiang_2010-Envelope_L1157,Tanner_2011-VLA_HH211}.
The right panel of Figure~\ref{fig-vla-temp} presents the derived kinetic 
temperature obtained from the simultaneous \ammo(1,1) and (2,2) line fit, 
with uncertainties in the temperature determination between 0.2\,K, in the central region, 
up to 0.5\,K in the outer regions. 
Surprisingly, the kinetic temperature map is quite constant, in particular, there 
is no evidence for an increase in temperature towards the peak continuum position.

\subsection{SMA}
\begin{figure}
\plotone{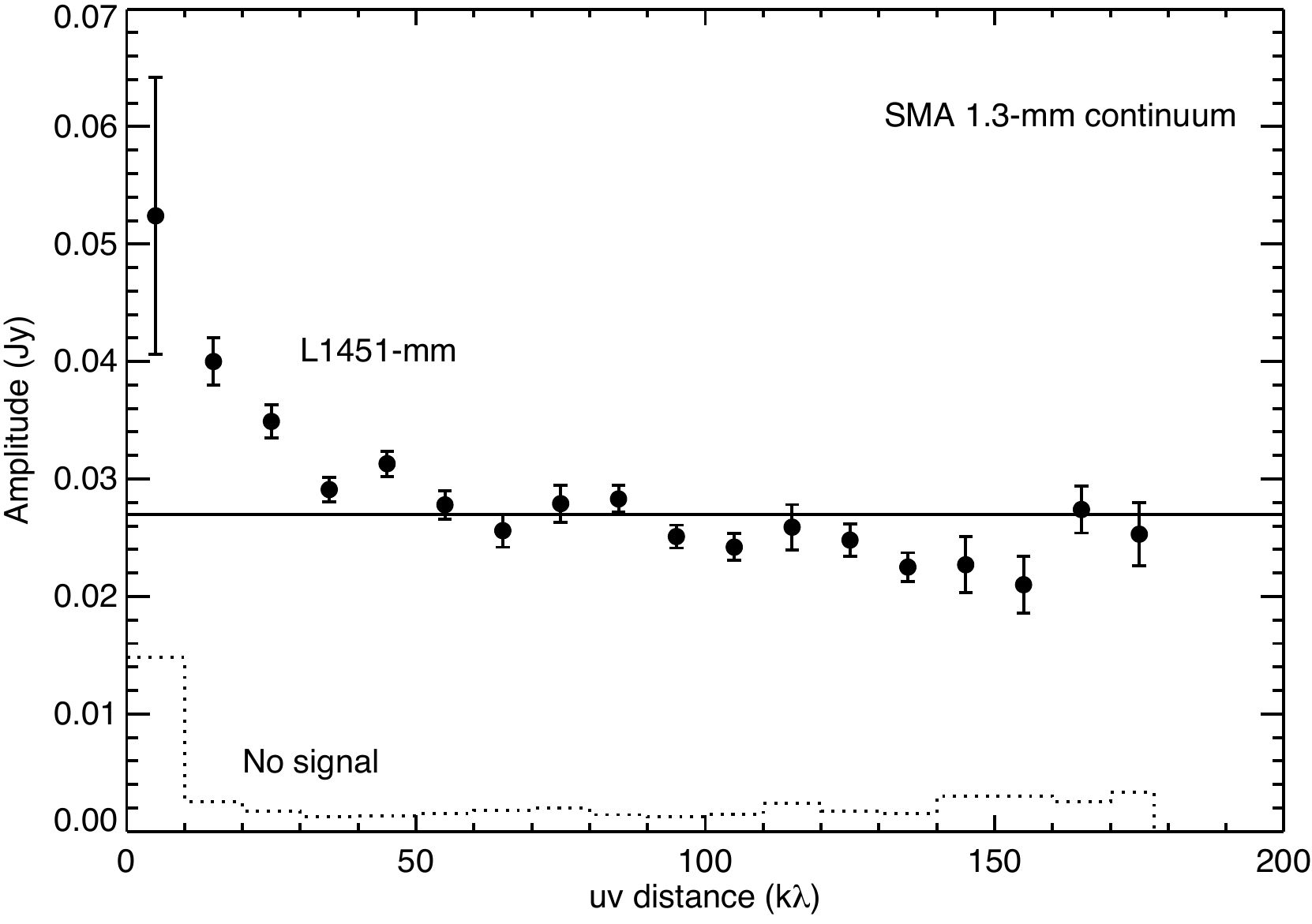}
\caption{Visibilities amplitude as a function of $uv$-distance. 
The dotted histogram indicates the expected amplitude in the absence of signal. 
The solid line shows the flux from the point source fit reported in Table~\ref{tab:fit}. 
\label{fig-uv-amp}}
\end{figure}

\begin{deluxetable*}{lcccccc}
\tablecolumns{7}
\tablecaption{Results of {\it uv} Fits of SMA 1.3-mm continuum data for L1451-mm
\label{tab:fit}}
\tablewidth{0pt}
\tablehead{
\colhead{}              & \multicolumn{2}{c}{Phase Center\tablenotemark{a}} & \colhead{Offset} & \colhead{Peak Flux} & \colhead{Size (FWHM)} & \colhead{PA}\\
\colhead{Source} & \colhead{$\alpha$(J2000)} & \colhead{$\delta$(J2000)} & \colhead{(arcsec)}  & \colhead{(mJy)} & \colhead{(arcsec)} & \colhead{($\deg$)}
}
\startdata
Point Source (vis. longer than 40$k\lambda$) & 3:25:10.21 & +30:23:55.3 & (0.40, $-$0.23)    & 27.0$\pm$0.4 & \nodata & \nodata \\
Gaussian (all vis.)      & \nodata & \nodata  & (0.40, $-$0.23) & 32.8$\pm$0.6 & (0.66$\pm$0.05, 0.45$\pm$0.05) & -88$\pm$8
\enddata
\tablenotetext{a}{Units of R.A. are hours, minutes, and seconds. 
Units of declination are degrees, arcminutes, and arcseconds.}
\end{deluxetable*}

The visibility amplitude as a function of $uv$-distance for the 1.3-mm continuum is shown 
in Fig.~\ref{fig-uv-amp}. 
From this figure we identify two components. An extended component, 
that is resolved at long baselines, and a compact component that remains 
unresolved even at the longest baselines (indicated by the horizontal 
line in Figure~\ref{fig-uv-amp}).
This unresolved component is commonly seen towards dense cores 
containing a central protostar, and it is interpreted as 
arising from an unresolved central disk 
\citep[e.g.,][]{Jorgensen_2007-PROSAC_I,Jorgensen_2009-PROSAC_II}.
However, in the case of \source there is no infrared detection of a central protostar.

\begin{figure}
\plotone{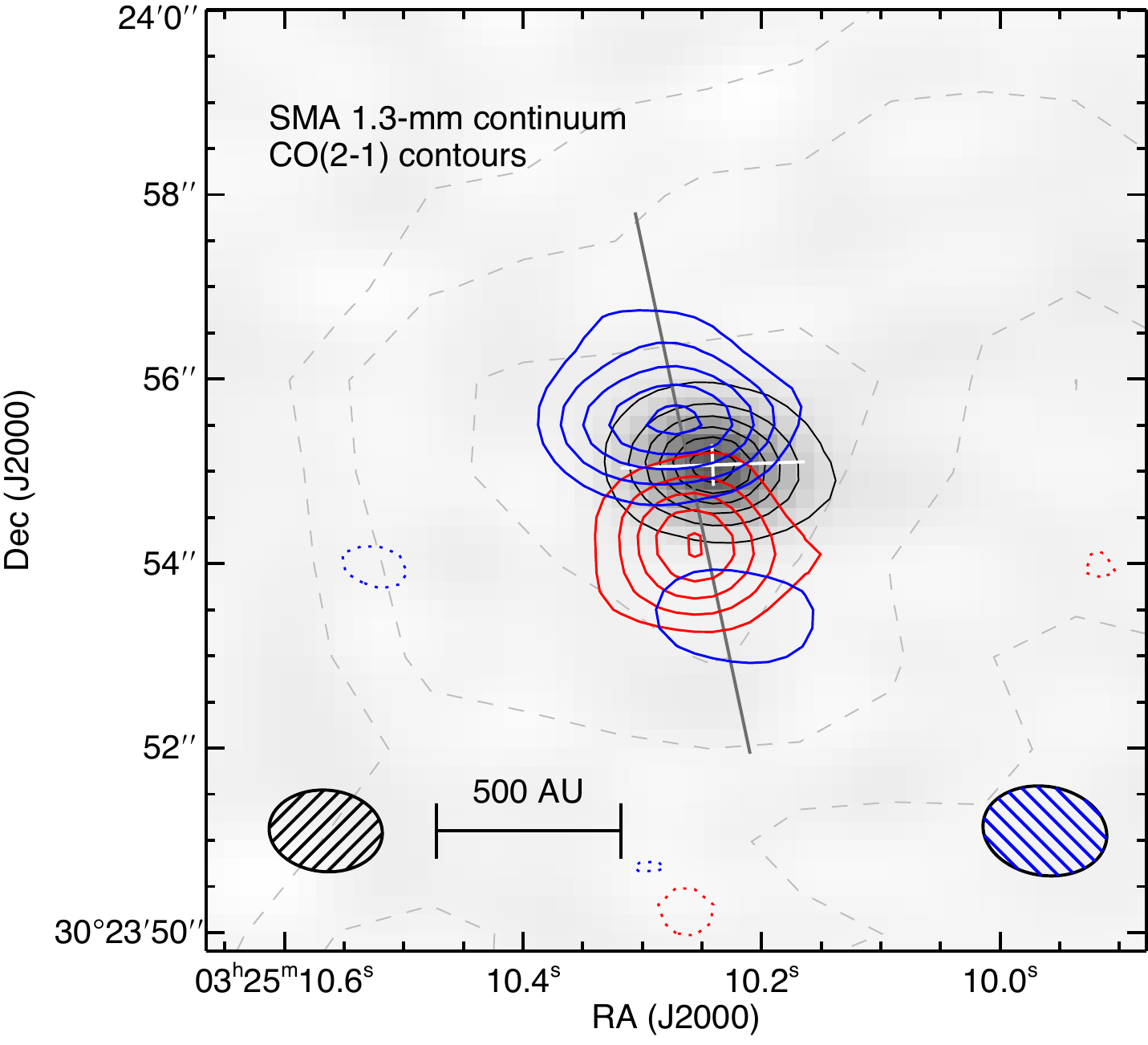}
\caption{The greyscale map shows the source detected in the 1.3-mm continuum continuum observed 
with SMA, with the overlaid black contours at [3,7,11,15,19,23]\,m\jybm. 
Overlaid are the contours for the \co(2--1) integrated intensity using the red and 
blue channels (red and blue channels are taken between 
5.3-6.9\,\kms and 1.9-3.7\,\kms, respectively). 
Dotted contours denote negative contour levels.
Contour levels are drawn at integer multiples of 114\,m\jybm\,\kms. 
Dashed light grey contours show the CARMA 3-mm continuum emission 
presented in Figure~\ref{fig_mambo_carma}.  
The white line shows the direction of the Gaussian fit on the uv-plane as reported in Table~\ref{tab:fit}. 
The gray line is cut for the position-velocity diagram shown in Figure~\ref{fig:pv_cut}. 
The 1.3-mm continuum \co(2--1) emission synthesized beams are shown at 
bottom left and right corners, respectively.
\label{fig-sma-map}}
\end{figure}

\begin{figure}
\plotone{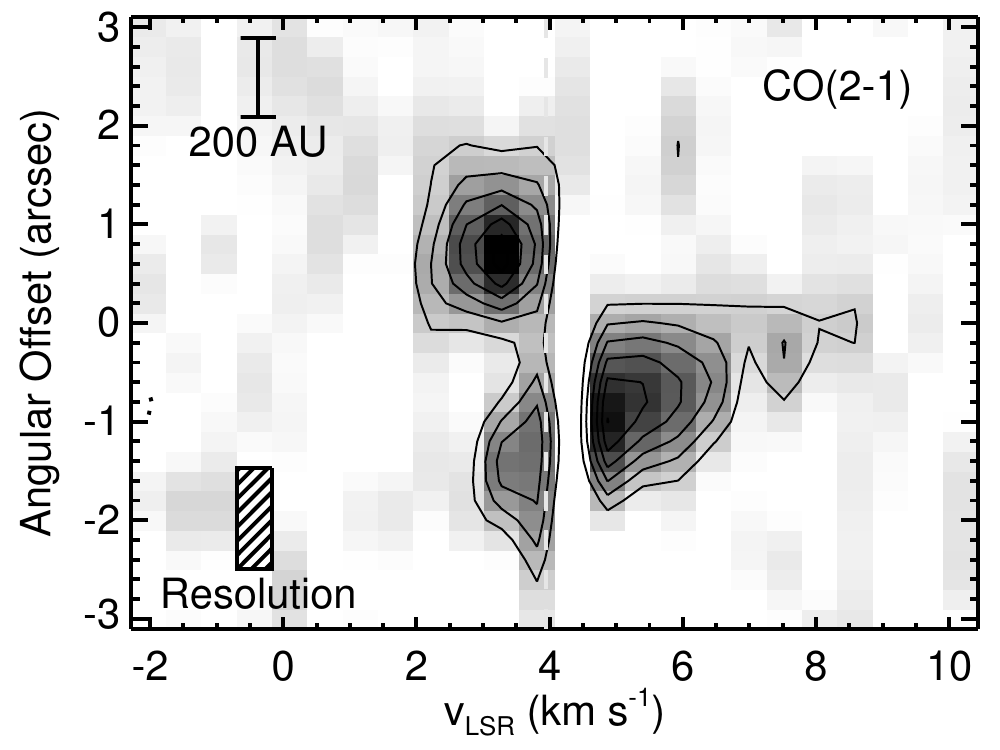}
\caption{Position velocity diagram for the SMA \co(2--1) data along the gray line shown 
in Figure~\ref{fig-sma-map}. 
Contours drawn start at 0.3\,\jybm with an increment of 0.24\,\jybm. 
The dashed vertical line shows the centroid velocity of the dense core, 3.94\kms.
\label{fig:pv_cut}}
\end{figure}

The SMA 1.3-mm continuum map is shown in Figure~\ref{fig-sma-map}, with red- 
and blue-shifted \co(2--1) emission overlaid. 
The dust continuum emission clearly shows the central source, 
also presented in Figure~\ref{fig-uv-amp}. 
The position of this continuum source coincides with the pixel where the CARMA 3-mm 
continuum peaks. 
The orientation of the SMA continuum emission, obtained through a fit of the 
visibilities and listed in Table~\ref{tab:fit}, 
is close to the right ascension axis and clearly different from the red- and blue-shifted 
\co emission.  
The detected \co(2--1) emission shows a large velocity dispersion, with spatially 
separated blue- and red-shifted lobes (see Figure~\ref{fig-sma-map}). 
Figure~\ref{fig:pv_cut} shows the Position Velocity (PV) diagram along the gray line 
drawn in Figure~\ref{fig-sma-map}, where the dashed vertical line shows the centroid velocity 
of the dense core, 3.94\,\kms, which is consistent with the interferometric observations 
(see Table~\ref{table-grad}) and the \ammo(1,1) data obtained with the GBT at 
30$\arcsec$ (Pineda et al. 2011, in prep).

The dust mass of the compact (unresolved) emission is estimated as: 
\begin{equation}
M_{1.3~mm} = 1.3~\msun \left(\frac{F_{1.3~mm}}{1{\rm Jy}}\right) \left(\frac{d}{200~{\rm pc}}\right)^{2}
\left( e^{0.36\,(30~K/T)} -1\right)~,
\end{equation}
where we assumed optically thin emission and a dust opacity per dust mass ($\kappa_{1.3~mm}$) 
of 0.86\,cm$^{2}$\,g$^{-1}$ \citep[thick ice mantles coagulated at $10^{5}$\,cm$^{-3}$ from][]{OH94} 
and a gas-to-dust ratio of 100, 
see \cite{Jorgensen_2007-PROSAC_I}. 
The flux from the unresolved emission is estimated by fitting a point source to 
baselines longer than 40~$k\lambda$, as in \cite{Jorgensen_2007-PROSAC_I}, and 
therefore it avoids contamination from the dense core itself. 
The result of fitting a point source gives a flux of $27.0$\,mJy, see Table~\ref{tab:fit}, 
which implies a mass of 
\begin{equation}\label{eq-Mdisk}
M_{1.3~mm}=0.024\,\msun
\end{equation}
for a temperature of 30\,K \citep[as used in][]{Jorgensen_2007-PROSAC_I}. 
This mass will be used as a first estimate for the circumstellar disk mass  
\citep[see][for a discussion]{Jorgensen_2007-PROSAC_I}.

The disk--to--dense core mass ratio, $M_{disk}/M_{dense~core}$, is estimated using the 
disk mass from eq.~\ref{eq-Mdisk} and the dense core mass. 
This ratio is low ($\approx$0.1) but comparable to Class~0 objects 
\citep{Enoch_2011-Serpens_Class0,Jorgensen_2009-PROSAC_II,Enoch_2009-Class0-Disk}.

\subsection{Simultaneous Fit of Visibilities and Broadband SED}
A powerful way to constrain the physical parameters of dense cores and YSOs 
is by fitting the broadband SED \citep[e.g.,][]{Robitaille_2007-SED_Fitter}. 
In the case of \source, only detections at 160 and 1200\,$\mu$m are 
available, which makes the broadband SED fit a not well-constrained problem. 
Here, the information from the 1.3-mm continuum observations (SMA) 
is extremely important to help discriminate between different physical models
\citep[e.g.,][]{Enoch_2009-Class0-Disk}.

In order to compare the continuum emission model with the interferometric 
observations, the model is sampled in uv-space to match the observations 
using the \verb+uvmodel+ task in MIRIAD.
The synthetic and observed visibilities are both binned in uv-distance (using 
the \verb+uvamp+ task in MIRIAD), and then 
they are added as an extra term to the $\chi^2$ to minimize,
\begin{equation}
\chi^2_{vis}=\sum_{i}\left( \frac{V_{i,obs} - V_{i,model}}{\sigma_{V_{i,obs}}} \right)^2, 
\end{equation}
where $V_{i,obs}$ and $V_{i,model}$ are the average observed and synthetic 
visibilities in the $i$ bin, respectively, while $\sigma_{V_{i,obs}}$ is the uncertainty 
of the observed average visibility.
The $\chi^2$ subject to minimization is 
\begin{equation}
\chi^2 = (1-Q) \chi^2_{SED} + Q \chi^2_{vis}, 
\end{equation}
where $Q$ is an ad-hoc weight used to 
control how important it is to fit the visibilities compared to the broadband SED. 
In this case a $Q=0.5$ is used, which gives the same weight to the SED and 
visibilities fit.

Because of the problem's high dimensionality, the $\chi^2$ minimization is 
carried out with a genetic algorithm \citep{Johnston_2011-RT_Model_Massive_Disk}, 
while the model 
SEDs and visibilities are calculated with a new Monte-Carlo radiation transfer code  
(Robitaille et al., 2011 in prep), which is based on the radiative transfer code
presented by \cite{Whitney_2003-2D_RT_Geometry}.  The new code uses raytracing for
the thermal emission at sub-mm and mm wavelengths, providing excellent
signal-to-noise to fit the long-wavelength SED and visibilities.

Using this fitting program we explore three models with increasing levels of 
complexity to explain our observations: 
a) starless isothermal dense core; 
b) dense core with a YSO and disk at the center; and 
c) dense core with a central FHSC.

\begin{figure}
\plotone{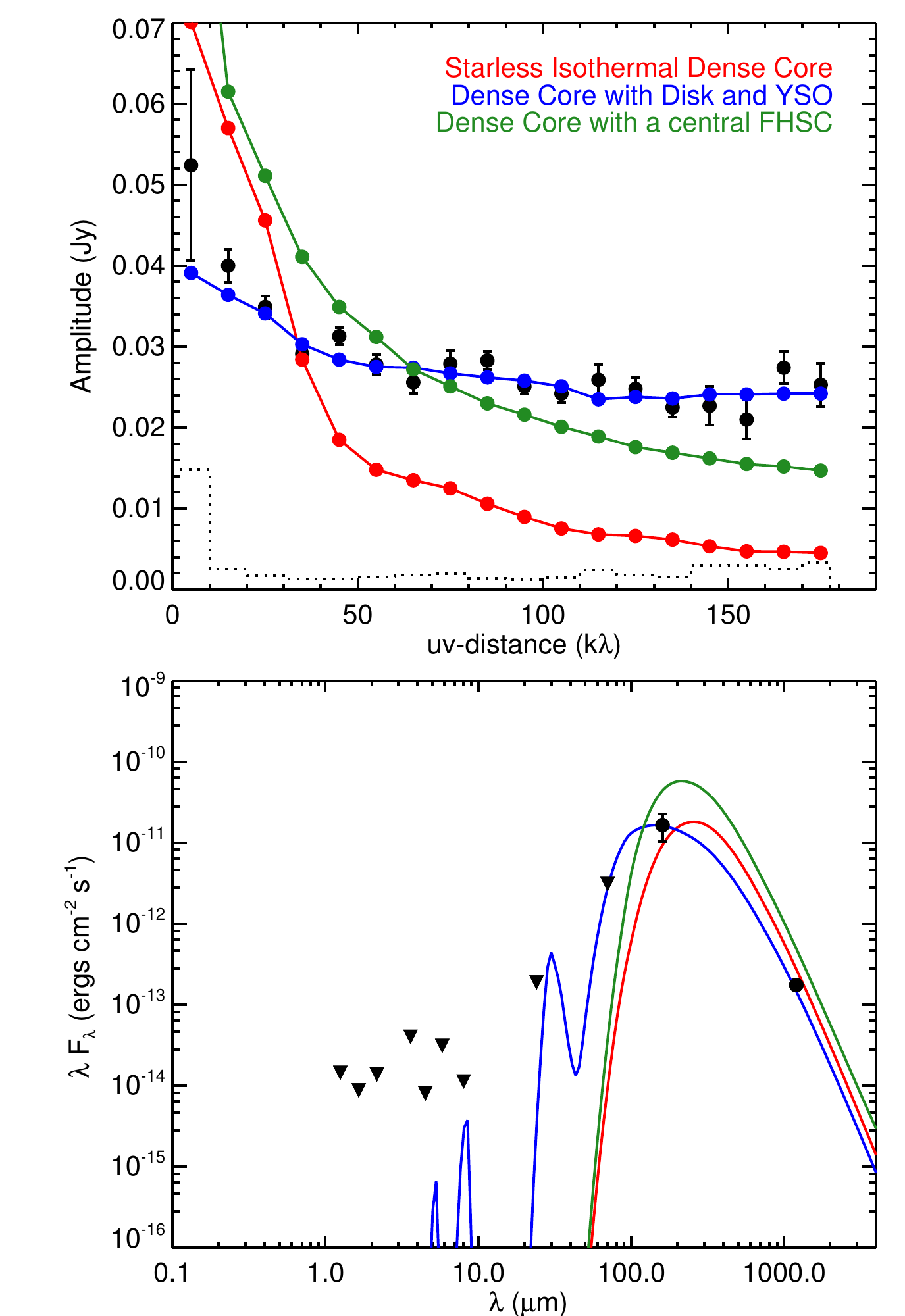}
\caption{%
Summary of best-fit of the broadband SED and dust continuum visibilities 
for three different models: 
starless isothermal dense core in red; 
dense core with a YSO and disk at the center in blue; and 
dense core with a central FHSC in green. 
The data are shown in black. 
The top panel shows the visibilities in filled circles.
The bottom panel shows the broadband SED for \source, where the upper 
limits are shown by triangles, measurements are shown by 
filled black circles, and the best model fits are shown by the solid curves.
\label{fig-sma-sed-all}}
\end{figure}

The parameter ranges searched using the genetic algorithm is given in Table~\ref{table:par-range}, 
a summary of the fit results is shown in Figure~\ref{fig-sma-sed-all}, 
and the model parameters are listed in Table~\ref{table:sed_model}.

\begin{deluxetable}{lccc}
\tablecolumns{4}
\tablewidth{0pt}
\tablecaption{Parameter Ranges Searched\label{table:par-range}
}
\tablehead{%
\colhead{Parameter} & 
\colhead{Description} & 
\colhead{Value/Range} & 
\colhead{Sampling}
}
\startdata
\cutinhead{Starless Dense Core}
$p$ & Exponent of density profile  &    $-2$  -- $0$  &    linear\\
$M_{dense~core}$ & Envelope Mass (\msun) &  $10^{-5}$ --  $10$ &       logarithmic\\
$T_{dust}$ & Envelope temperature (K)  &   $5$ --  $30$ & linear\\
$R_{max}$ &  Envelope outer radius (AU) &  $1,000$ -- $5,000$ &       logarithmic\\
\cutinhead{Dense Core with Disk and YSO}
$L_{int}$        & Intrinsic luminosity (\lsun) & $10^{-4}$ -- $0.1$ &        logarithmic\\
$\dot{M}_{infall}$ &  Infall rate (\msunyr) &  $10^{-9}$ -- $10^{-6}$ &  logarithmic\\
$R_{env}$ &  Outer envelope radius  (AU)  &   $1,000$ -- $5,000$  &  logarithmic\\
$R_{cent}$ &  centrifugal radius (AU)  &      $50$ --   $1,000$    &  logarithmic\\
$R_{disk}$ &  outer disk radius (AU) & $50$ --  $1,000$  &  logarithmic\\
$M_{disk}$ & Disk Mass (\msun) &  $10^{-8}$ -- $10^{-3}$   &  logarithmic\\
$i$ & Viewing Angle ($\deg$) &   $0$ -- $90$ & linear
\enddata
\end{deluxetable}

\paragraph{Starless isothermal dense core}
The simplest model consist of a pure isothermal dense core with a density 
profile $n \propto r^{-p},$ where the exponent, $p$, of the density profile is 
a free parameter, but constrained to values smaller than $2$. 
The dense core temperature, $T_{dust}$, and the outer radius, $R_{env}$, are also 
parameters in the fit, which are listed in Table~\ref{table:sed_model} along 
with the dense core mass, $M_{env}$. 
The best model fit for the starless isothermal dense core 
model is shown in red in Figure~\ref{fig-sma-sed-all}, where it shows that this model 
does not provide a good match to the visibilities. If the power-law 
density exponent is not constrained, then a very steep density profile, $n\propto r^{-2.8}$,
can actually match both the SED and visibilities. 
However, a power-law exponent of $-2.8$ is beyond the range deemed 
physically reasonable.
Because, even though the outer region of starless cores (and cylindrical filaments) 
can have similar steep density profiles \citep[e.g.,][]{Tafalla_2002-Freeze},  
in the inner region ($r<$3,000~AU) their density profiles are flat, 
which is exactly the region we are interested in to produce the compact emission.

\paragraph{Dense Core with a central YSO and disk}
The next model fitted is one composed of a dense core with a central 
YSO and disk. 
The dense core is modelled as a rotating and infalling envelope \citep{Ulrich_1976-infall_model}, 
with outer radius $R_{\rm env}$, total mass $M_{\rm env}$,  and infall rate $\dot{M}_{\rm infall}$. 
The density of the envelope is given in spherical polar coordinates by
\begin{equation}
\rho(r,\theta,\phi) = \rho_0^{\rm env}\left(\frac{r}{\rc}\right)^{-3/2}\left(1 + \frac{\mu}{\mu_0}\right)^{-1/2}\left(\frac{\mu}{\mu_0} + \frac{2\mu_0^2\rc}{r}\right)^{-1}
\end{equation}
where $\rc$ is the centrifugal radius, $\mu=\cos{\theta}$, and $\mu_0$ is the cosine of the polar angle of a streamline of infalling particles as $r\rightarrow\infty$, which is given by:
\begin{equation}
\mu_0^3 + \mu_0\left(\frac{r}{\rc} - 1\right) - \mu\left(\frac{r}{\rc}\right) = 0
\end{equation}
The normalization constant $\rho_0^{\rm env}$ is related to the infall rate by: 
\begin{equation}
\rho_0^{\rm env} = \frac{\dot{M}_{\rm infall}}{4\pi\left(G\mstar\rc^3\right)^{1/2}}~,
\end{equation}
where $\mstar$ is the mass of the central object.
The disk is modeled as a passive flared disk described in cylindrical polar coordinates by
\begin{equation}
\rho(R,z,\phi) = \rho_0^{\rm disk}\,\left(\frac{100\,AU}{R}\right)^{\beta - q}\,\exp{\left[-\frac{1}{2}\left(\frac{z}{h(R)}\right)^2\right]}
\end{equation}
where $\rho_0^{\rm disk}$ is defined by the disk mass $\mdisk$, $q$ is the surface density radial exponent (which we set to $-1$), $\beta$ is the disk flaring power (set to $1.25$), and the disk scale-height $h(R)$ is given by:
\begin{equation}
h(R) = h_0\,\left(\frac{R}{100\,AU}\right)^\beta.
\end{equation}
where $h_0$ is the scale-height at 100\,AU and it is set to 10~AU.

The viewing angle $i$ is a parameter in the fitting.
The central protostar is modeled as an object with an effective surface temperature of 3,000\,K and 
intrinsic luminosity $L_{int}$, where the parameter $L_{int}$ includes the luminosity due to accretion.

The temperature is computed self-consistently with the density using
the radiation transfer code, 
see \cite{Whitney_2003-2D_RT_Geometry,Whitney_2003-2D_RT_Evolution,Whitney_2004-2D_RT_Temp}. 
It assumes a geometry (e.g., Ulrich envelope model with a flared disk), the dust properties, 
and local thermodynamic equilibrium. 
Here we use dust opacities of \cite{OH94} for dust grains with thick ice mantles after $10^{5}$\,years 
of coagulation at a density of $10^{6}$\,cm$^{-3}$.

The best model parameters are listed in Table~\ref{table:sed_model}.  
This model provides an excellent fit to the visibilities, while the SED fit underestimates the flux at 
$160\mu m$. 

\begin{deluxetable*}{lcccccccc}
\tablewidth{0pt}
\tablecaption{Best Model Fit Parameters
\label{table:sed_model}}
\tablehead{ %
\colhead{Model} & \colhead{$L_{int}$} & \colhead{$\dot{M}_{infall}$\tablenotemark{a}} & \colhead{$M_{disk}$} &
\colhead{$R_{disk}$} & \colhead{Viewing Angle} & 
\colhead{$M_{dense~core}$} & \colhead{$R_{env}$} & \colhead{$T_{dust}$}\\
\colhead{} & \colhead{(\lsun)} & \colhead{(\msunyr)} & \colhead{(\msun)} & 
\colhead{(AU)} & \colhead{($\deg$)} & \colhead{(\msun)} & \colhead{(AU)} & \colhead{(K)}
}
\startdata
Starless Dense Core          & \nodata &  \nodata &  \nodata  &  \nodata  &  \nodata  & $0.353$ & $1001$ & $10.2$ \\
Dense Core with Disk and YSO &  $0.0450$ & $7.1\times 10^{-6}$ & $0.086$ & $107$ & $64.9$ & $0.146$ & $1007$ & \nodata \\
Dense Core with central FHSC & \nodata &  \nodata &  \nodata  &  \nodata  &  \nodata  & $0.56$ & $2543$ & \nodata

\enddata
\tablenotetext{a}{Infall rate derived using eq.~(\ref{eq:infall}) for $\mstar=0.086\,\msun$.}
\end{deluxetable*}

Given the best fit result we estimate the accretion luminosity, $L_{acc}$,
as
\begin{equation}
L_{acc} = \frac{G\,M_{*}\,\dot{M}_{acc}}{R_{*}}~, \label{eq:lacc}
\end{equation}
where $M_{*}$ and $R_{*}$ are the protostellar mass and radius, and 
$\dot{M}_{acc}$ is the accretion rate onto the protostar. 
Using equation~(\ref{eq:lacc}) we obtain an accretion luminosity of $6\, \lsun$, 
where we have assumed that the accretion rate is the same as the infall rate, 
\begin{equation} \label{eq:infall}
\dot{M}_{acc}=\dot{M}_{infall}= 2.4\times 10^{-5}\,\left(\frac{\mstar}{\msun}\right)^{0.5}\msunyr, 
\end{equation}
the minimum mass of the central object is the mass of the disk, 
$\mstar=M_{disk}=0.086\,\msun$, and 
that the central object might be a young protostar, $R_{*}=3\,\rsun$. 
This expected accretion luminosity is $100$ times higher 
than what can be kept hidden at the center of the core in the best-fit model, and 
therefore some of the assumptions must be clearly misrepresenting reality. 
Another simple estimate that is derived from equation~(\ref{eq:lacc}) is the accretion rate 
onto the central object needed to produce the same luminosity of the best model, 
obtaining $\dot{M}_{acc}=2.45\times10^{-8}\,\msunyr$ 
(much lower than the estimated infall rate, $7.1\times 10^{-6}\,\msunyr$).
Clearly, in order to make a self-consistent model an extremely low accretion rate must be assumed.

\paragraph{Dense Core with a central FHSC}
Another possibility to reconcile the low luminosity observed and the 
accretion rates expected from the models presented above is to increase the stellar radius, $R_*$. 
There is one object which has been predicted from numerical simulations 
\citep{Larson_1969-Collapse,Masunaga_1998-First_Collapse,Masunaga_2000-Second_Collapse,Machida_2008-Outflow_Core}
that is large enough to fit the description: \emph{the first hydrostatic core} (FHSC).

\cite{Masunaga_1998-First_Collapse} performed radiation hydrodynamic 
simulations of a collapsing core until the formation of a FHSC and found 
that for an initial core of 0.3\,\msun (model M3a), 
a FHSC of $R_*\sim 5\,$AU and $L_{int}\approx 0.03\,\lsun$  
(including accretion luminosity) is formed. 
We use the density and temperature profile resulting from this simulation as the input 
parameter for the radiative transfer code. 
The mass and radius of the dense core are varied (see Table~\ref{table:sed_model}) 
to find the best match of 
the observed SED and visibilities, see green points in Figure~\ref{fig-sma-sed-all}. 
Notice that this model does not use the same density and temperature profile as used in 
the dense core with disk and YSO model, and this explains the different SEDs.
It is clear that the FHSC model visibilities do not provide as good a match to the observations as 
the YSO model, however, the FHSC model only have two free parameters compared to the 
seven free parameters of the YSO model.

\subsection{The nature of the \co(2--1) emission}\label{sec-outflow-explain}
If there is no source of heating within the core, then \co should freeze-out onto dust 
grains \citep[e.g.,][]{Tafalla_2002-Freeze}. 
Therefore, the presence of the \co(2--1) emission in itself is strongly suggestive of a 
central heating source. 
From the \ammo, \nh and \dammo centroid velocity maps it is clear that the  
velocity gradient is in the RA direction while the \co emission is more or less 
perpendicular. 
Since the orientation of the dust continuum emission detected with SMA is almost 
perpendicular to the \co(2--1) emission, we argue in favor of a central object and outflow system.

The amount of mass needed to keep material at 560\,AU with a velocity of 1.3\,\kms 
(similar parameters to the \co emission) is $\approx 0.53\,\msun$, which is almost twice 
the total mass in the dense core. 
Therefore, despite the low velocity seen in the \co(2-1) emission this gas is unbound  
and it is consistent with \co(2--1) tracing a slow molecular outflow.

\subsection{Outflow Properties}\label{sec-outflow-properties}
The physical parameters of molecular outflows are typically calculated using 
both \co and \thco lines \citep[e.g.,][]{Arce_2010-outflow_Perseus}.
Unfortunately, in \source there are no detections of \thco(2--1) and we are left to use only
the \co(2--1) emission to study the high-velocity gas.

A lower limit for the mass entrained by the outflow, $M_{flow}$, is estimated assuming 
that the \co(2--1) emission is optically thin (see details in Appendix~\ref{sec:Nco}) and with  
an excitation temperature of $T_{ex}=20$~K.
The momentum ($P_{flow}$) and energy ($E_{flow}$) of the outflow along the line of sight 
are estimated following 
\cite{CabritBertout_1990-Outflow_Properties},
\begin{eqnarray}
P_{flow} &=&  \sum_i M_{out,i}\, |v_i -v_{center}| ~,\\
E_{flow} &=&  \frac{1}{2} \sum_i  M_{out,i} \, |v_i -v_{center}|^2~,
\end{eqnarray}
where $M_{out,i}$ is the mass in voxel $i$, $v_{center}$ is the velocity of the 
core, and $v_{i}$ is the velocity of voxel $i$. 
Also, the outflow characteristic velocity, $v_{flow}$, is calculated as $P_{flow}/M_{flow}$.
An upper limit for the lobe size, $R_{lobe}$, is estimated from the (not deconvolved) 
extension of the red- and blue-shifted 
\co emission ($2\,R_{lobe}$) $\approx 4.5\arcsec$, or $\approx 1,120$~AU at 
the distance of Perseus. 
We also calculate the dynamical time, $\tau_{dyn}=R_{lobe}/v_{flow}$, 
mechanical luminosity, $L_{flow}=E_{flow}/\tau_{dyn}$, 
force, $F_{flow}=P_{flow}/\tau_{dyn}$, and rate, $\dot{M}_{flow}=M_{flow}/\tau_{dyn}$.

The outflow properties are calculated using only the voxels with signal-to-noise ratio 
higher than $2$.
To avoid contamination in the outflow parameters from the cloud emission 
the central channel is not used in the calculations, 
and a second estimate is calculated where the three central channels are removed.
The outflow parameters are reported in Table~\ref{table:outflow}, where we notice that 
the differences between the quantities calculated using both methods are small.

The outflow properties presented in Table~\ref{table:outflow} show an extremely 
weak outflow in \source. However, when compared to the outflow found towards L1014 
by \cite{Bourke_2005-outflow_L1014} both $P_{flow}$ and $E_{flow}$ are 
close to the lower limits estimated using similar assumptions.

\begin{deluxetable}{lc}[h]
\tablewidth{0pt}
\tablecaption{Summary of Outflow Properties\tablenotemark{a}\label{table:outflow}}
\tablehead{
\colhead{Property} & 
}
\startdata
Mass (\msun)    			& $1.2\times 10^{-5}$	($8.4\times 10^{-6}$)\\
Momentum (\msun \kms) 		&  $1.7\times 10^{-5}$	($1.4\times 10^{-5}$)\\
Energy (ergs) 				& $3.1\times 10^{38}$ 	($3.0\times 10^{38}$)\\
Luminosity (\lsun) 			&  $1.3\times 10^{-6}$	($1.6\times 10^{-6}$)\\
Force  (\msun \kms yr$^{-1}$) 	& $8.3\times 10^{-9}$   	($9.3\times 10^{-9}$)\\
Characteristic Velocity  (\kms) 	&  $1.3$  				($1.7$)\\
Dynamical Time (yr)  		&  $2.0\times 10^{3}$	($1.6\times 10^{3}$)\\
Outflow rate (\msunyr) 		& $6\times 10^{-9}$ 		($5\times 10^{-9}$)
\enddata
\tablecomments{Properties are calculated without the central channel, 
while the values in parentheses are calculated without the 3 central channels.}
\tablenotetext{a}{Not corrected for outflow inclination with respect to the plane of the sky.}
\end{deluxetable}

The dynamical time is consistent with the FHSC estimated lifetime 
\citep[e.g.,][]{Machida_2008-Outflow_Core}. 

Recent three-dimensional radiation magneto hydrodynamics (RMHD) simulations 
of the dense core collapse 
\citep{Machida_2008-Outflow_Core,Tomida_2010-First_Core_RMHD,Commercon_2010-MHD_radiation} 
show that when the FHSC is formed a slow outflow can be driven 
even before the existence of a protostar 
\citep[see also][]{Tomisaka_2002-Core_Collapse,Banerjee_2006-Outflow_Cores}. 
In these simulations, the outflow that is generated is poorly collimated and 
typically has maximum velocity of 3\,\kms -- very similar to the observed outflow in \source.
In contrast, theoretical models 
\citep{Shang_2007-PPV_Outflow_Theory,Pudritz_2007-PPV_DiskWind} 
and observations \citep[e.g.,][]{Arce_2007-PPV} indicate that outflows from young 
protostars are highly collimated and exhibit velocities of a few tens of \kms
although outflows from VeLLOs display lower velocities 
\citep{Andre_1999-Discovery_IRAM04191,Bourke_2005-outflow_L1014}. 

The properties of the \source molecular outflow are consistent with a picture where 
a first hydrostatic core is the driving source of the poorly collimated and slow outflow observed.

\section{Discussion}\label{sec:discussion}

The detection of an unresolved source of continuum in the CARMA and SMA observations  
strongly suggests the presence of a central source of radiation and/or a disk. 
Moreover, the simultaneous fit of the broadband SED and the continuum visibilities rules out 
the possibility of explaining the observations without a central source (either a YSO or a FHSC). 
From the SED modeling it appears feasible to hide the central YSO even at 70$\mu$m, 
but it also requires an extremely inefficient or episodic accretion process. 
This might be consistent with the results obtained by 
\cite{Enoch_2009-Properties_YSOs,Dunham_2010-YSO_evol} and \cite{Offner_2011-YSO_luminosity}, 
where episodic accretion is argued to explain the low luminosity of YSOs observed 
by Spitzer.

If the unresolved emission observed with the SMA is interpreted as a disk 
\citep[as done by][]{Jorgensen_2007-PROSAC_I}, then 
the disk mass is already $\sim$10\% of the dense core mass, which 
is similar to the disk mass found in class 0 objects 
\citep[e.g.,][]{Enoch_2011-Serpens_Class0,Enoch_2009-Class0-Disk,Maury_2010-PdBI_pilot_survey}, 
although \cite{Belloche_2002-Molecules_IRAM04191_PdBI} shows evidence for a small disk 
in the young protostar IRAM~04191 ($M_{disk}<10^{-3}\msun$ and $M_{core}\approx 1.5\,\msun$).
These studies of Class~0 objects suggests that the assembly of mass to form a disk starts very early on.

A slow molecular outflow is detected in the \co(2--1) line, see Sec.~\ref{sec-outflow-explain} 
and \ref{sec-outflow-properties}. 
Its orientation is almost perpendicular to the velocity gradient seen in dense gas tracers 
observed with VLA and CARMA (\ammo, \nh and \dammo), and despite the low velocity 
the gas is unbound. 
The properties presented in Table~\ref{table:outflow} place it as the weakest outflow found so far, 
with the lowest energy and momentum measured. 
Unfortunately, we have no estimate of the outflow inclination angle and therefore some of the  
outflow parameters might be underestimated. 
If the outflow is close to the plane of the sky, then the outflow velocity would be faster but 
it still could be consistent with a slow outflow driven by a FHSC depending on how large is 
the correction. This, however, would imply that the outflow extension is the one measured 
in the data, and therefore the outflow would have a shorter dynamical time and low degree 
of collimation. 
On the other hand, if the outflow is nearly in the line-of-sight, then the outflow velocity is similar 
to the 1.5\,\kms measured from the data. 
However, the outflow extension would be much larger implying a longer dynamical time, 
which might be similar to those predicted by \cite{Tomida_2010-FHSC_Model} for FHSCs 
in recent numerical simulations. 
Therefore, constraining the inclination angle 
\citep[e.g., through observations of the outflow cavity as in][]{L1014-Tracy} 
would provide important insight regarding the 
outflow and by extension to the central object.

For all the reasons listed above, we claim that a central source of radiation 
(either a YSO or a FHSC) must be present within \source. 
The lack of sensitive observations at mid-infrared wavelengths 
restricts our ability to carry out a more detailed modeling of this object.
From our best-fit models we predict that \source should be detected by 
the Herschel Gould Belt Survey \citep{Andre_2005-Herschel_Gould_Belt}, 
similar to the observations by \cite{Linz_2010-UYSO_Herschel}. 
Therefore, those observations will provide a definitive answer regarding the  
luminosity of the central source and give more constraints to the modeling. 

One way to explain our observations is by having a first hydrostatic core at 
the center of the dense \source core, instead of a YSO.  
The simultaneous fit of both visibilities and broadband SED shows that 
a FHSC can also provide a good fit to the observations, 
with the advantage of having an accretion 
luminosity consistent with the observations. 
The presence of a slow and poorly collimated outflow further supports this scenario. 
It is for these reasons that we propose \source to be a FHSC candidate.

Future observations of \source with interferometers using a more extended 
configuration and/or different frequencies will probe the currently unresolved 
continuum emission. 
We expect that such observations will provide a constraint on the origin 
of the emission (i.e., disk or first hydrostatic core).
And, if the disk is confirmed, then a comparison with more evolved disks can be carried out. 
Moreover, observations of \co(3--2) would provide an estimate of the gas temperature, 
and therefore a good test to confirm that the \co emission is generated by an outflow 
(where the gas is usually warm). 
 
It is very important to note that three out of the four known FHSC candidates are found in the same 
molecular cloud \citep{Enoch_2010-FHSC_Perseus,Chen_2010-First_Core}. 
We compare this number to the expected number of FHSC in Perseus assuming a constant star formation rate, 
which can be estimated as 
\begin{equation}
N_{FHSC}=N_{Class~0}\frac{\tau_{FHSC}}{\tau_{Class~0}}~.
\end{equation}
We estimate the FHSC lifetime to be $\sim10^3$\,yr  
\citep[e.g.,][]{Machida_2008-Outflow_Core},  
the number of Class~0 sources in Perseus is $20-35$ 
\citep{Hatchell_2007-SED,Enoch_2009-Properties_YSOs}, and the 
Class~0 lifetime is $2-5\times 10^5$\,yr 
\citep{Visser_2001-SCUBA_lifetimes,Hatchell_2007-SED,Enoch_2009-Properties_YSOs}. 
Finally, the expected number of FHSC in Perseus is $\le 0.2$ objects \citep[similar results are 
obtained using statistics for Class~I objects, e.g.,][]{Evans_2009-c2d_summary}, and 
therefore, if all three candidates are confirmed, either Perseus is in an 
extremely peculiar epoch (e.g., a recent burst on the star formation rate) or this stage is 
longer than previously predicted by numerical simulations.
A longer lifetime for the FHSC stage, up to $10^{4}$\,yrs, has recently been suggested 
by \cite{Tomida_2010-FHSC_Model} for FHSCs formed in low-mass dense cores 
($\sim0.1$\,\msun). 

\section{Summary}
We present IRAM 30-m, CARMA, VLA, and SMA observations of the isolated 
low-mass dense core \source in the Perseus Molecular Cloud. 
No point source is detected towards the center of the core in NIR and Spitzer 
observations; however, a dust continuum source is identified in both CARMA and 
SMA continuum maps. 
Upper limits on the bolometric luminosity and temperature, $L_{bol}$ and 
$T_{bol}$, of 0.05\,\lsun and $30$\,K are estimated. 
Also, \co(2--1) emission is observed towards \source suggestive of a 
slow and poorly collimated outflow. 
Modeling the broadband SED and observed visibilities at 1.3-mm confirms the 
need for a YSO or a First Hydrostatic Core (FHSC) to explain the observations. 
However, more high-resolution observations are needed to distinguish between 
these two scenarios.

Although YSO and FHSC models are almost indistinguishable, the FHSC scenario 
seems more likely from the data at hand (and thus we may call \source a FHSC 
candidate).

Finally, if all current FHSC candidates are confirmed, 
then an important revision of the FHSC lifetime must be carried out, which may include 
modifications of the numerical simulations  \citep[e.g.,][]{Tomida_2010-FHSC_Model}.

\acknowledgments
JEP acknowledges support by the NSF through grant \#AF002 from the Association of Universities for Research in 
Astronomy, Inc., under NSF cooperative agreement AST-9613615 and by Fundaci\'on Andes under project No. C-13442. 
Support for this work was provided by the NSF through awards GSSP06-0015 and GSSP08-0031 from the NRAO.
This material is based upon work supported by the National Science Foundation under Grants 
No. AST-0407172 and AST-0908159 to AAG and AST-0845619 to HGA. 
T.L.B. acknowledges support from NASA Origins grant NXX09AB89G. 
G.A. acknowledges support from MICINN AYA2008-06189-C03-01 grant (co-funded with FEDER funds), 
and from Junta de Andaluc\'{\i}a.

Facilities:  \facility{CARMA}, \facility{VLA}, \facility{SMA}, \facility{IRAM:30m}, \facility{Spitzer}

\appendix
\section{Calculation of CO column density}\label{sec:Nco}

If the levels of the molecule are populated following a Boltzmann distribution of temperature $T_{ex}$, 
then the column density can be expressed as,
\begin{eqnarray}
N_{J} &=& \frac{8\pi \nu^{3}}{c^{3}}\frac{g_{J}}{g_{J+1}} \frac{1}{A_{J+1\rightarrow J}}
                    \frac{\int \tau\, dv}{\left(1-e^{-h\nu/kT_{ex}}\right)} \nonumber \\
N_{J} &=& 93.28 \left(\frac{2J+1}{2J+3}\right) \frac{\nu^{3}}{A_{J+1\rightarrow J}}\frac{\int \tau\, dv}{\left(1-e^{-T_{0}/T_{ex}}\right)}\, \label{eq:NJtau}
\end{eqnarray}
where $g_{J}=(2J+1)$ is the statistical weight of level $J$ for a linear rotor molecule, 
$\nu$ is the transition frequency in units of GHz, 
$A_{J+1\rightarrow J}$ is the spontaneous emission coefficient in s$^{-1}$, 
$\tau$ is the transition optical depth, 
the velocity is in \kms, and 
\begin{equation}
T_{0}\equiv \frac{h\,\nu}{k}\,.
\end{equation}
In the case of \co(2--1), we use $\nu=230.538\,{\rm GHz}$ and $A_{2\rightarrow 1} = 6.91\times 10^{-7}\,{\rm s}^{-1}$ 
(obtained from Leiden Atomic and Molecular Database\footnote{\url{http://www.strw.leidenuniv.nl/$\sim$moldata/}}), and 
therefore $T_{0}=11.0641~{\rm K}$.

Using the equation of radiative transfer to relate the observed emission, $T_R$, with 
excitation temperature, $T_{ex}$, and background temperature, $T_{cmb}$, we obtain:
\begin{eqnarray}
T_{R} &=& T_{0} \left[ \frac{1}{(e^{T_{0}/T_{ex}}-1)} - \frac{1}{(e^{T_{0}/T_{cmb}}-1)} \right]\left(1-e^{-\tau}\right) \nonumber \\
\int T_{R}\, dv &=& T_{0} \left[ \left(e^{T_{0}/T_{ex}}-1\right)^{-1} - \left(e^{T_{0}/T_{cmb}}-1\right)^{-1} \right] \int \tau\, dv ~,\label{eq:taudv}
\end{eqnarray}
where optically thin emission is assumed. 

Combining equations (\ref{eq:NJtau}) and (\ref{eq:taudv}), the column density of the level $J$ can be calculated as,
\begin{eqnarray}
N_{J} &=& 93.28 \left(\frac{2J+1}{2J+3}\right) \frac{\nu^{3}}{A_{J+1\rightarrow J}}
                   \frac{\int T_{MB}\, dv}{\left(1-e^{-T_{0}/T_{ex}}\right)T_{0} 
                   \left[ \left(e^{T_{0}/T_{ex}}-1\right)^{-1} - \left(e^{T_{0}/T_{cmb}}-1\right)^{-1} \right]} \nonumber \\
N_{1}(\co) &=& 9.924\times10^{14} 
		\frac{\int T_{MB}\, dv}{\left(1-e^{-11.06/T_{ex}}\right)11.06 \left[ \left(e^{11.06/T_{ex}}-1\right)^{-1} - \left(e^{11.06/T_{cmb}}-1\right)^{-1} \right]} 
	\ {\rm cm^{-2}},  \label{eq:NCO-1}
\end{eqnarray}
where equation~(\ref{eq:NCO-1}) gives the column density of the level $J=1$ of \co  using the \co(2--1) transition emission.

The total column density of \co is calculated as 
\begin{equation}
N(\co) = \frac{Z}{g_{J} e^{-E_{J}/k\,T_{ex}}} N_{J}(\co)= \frac{Z}{(2J+1)}e^{J(J+1) B\, h/k\,T_{ex}} N_{J}(\co)~,
\end{equation}
where $B$ is the rotation constant for a linear rotor ($B=57.635968$\,GHz for \co), and 
$Z$ is the partition function, which can be approximated as $Z\approx k\,T_{ex}/(h\,B)$. 
Therefore, in the case of \co $J=1$  we obtain,
\begin{equation}
N(\co) =
1.2\times10^{14} \, \frac{T_{ex}/11.06 }{\left(e^{11.06/T_{ex}}-1\right)^{-1} - \left(e^{11.06/T_{cmb}}-1\right)^{-1}}
\frac{e^{5.53/T_{ex}} \,\int T_{MB}\, dv}{\left(1-e^{-11.06/T_{ex}}\right)} 
\ {\rm cm^{-2}},
\end{equation}
which when combined with a \co abundance with respect to \htw, 
$[\co/\htw]=10^{-4}$, provides an estimate of the total column density of \htw.

The final conversion between column density and mass is done using
\begin{equation}
M = 2.71 \times 10^{-8}
        \left(\frac{N(\co)}{10^{14}\,{\rm cm^{-2}}}\right) \left(\frac{[\htw/\co]}{10^{4}}\right) \left(\frac{d}{250\,{\rm pc}}\right)^{2} 
        \left(\frac{A_{sky}}{\rm arcsec^{2}}\right)  \left(\frac{\mu}{2.3}\right)  \msun\,,
\end{equation}
where $A_{sky}$ is the area on the sky used to calculate $N(\co)$, and $\mu$ is the mean molecular weight.

\end{document}